\documentclass[nofootinbib,preprintnumbers,amsmath,amssymb,aps,prc,showkeys]{revtex4-1}

\usepackage{graphics}
\usepackage{graphicx}
\usepackage{color}
\usepackage{hyperref}
\usepackage{bm}
\usepackage{calligra}
\usepackage[T1]{fontenc}
\usepackage{amsmath}
\usepackage{amssymb}
\usepackage{amsfonts}
\usepackage{mathrsfs}  
\usepackage{dsfont}
\usepackage{footmisc}

\newcommand{\trento}{T\raisebox{-.5ex}{R}ENTo }
\newcommand{\trentonosp}{T\raisebox{-.5ex}{R}ENTo }
\def\vfs{v_{\rm fs}}
\def\tfs{\tau_{\rm fs}}

\begin{document}

\title{Pre-hydrodynamic evolution in large and small systems}

\author{T. Nunes da Silva}
\email{t.j.nunes@ufsc.br}
\affiliation{%
 Departamento de F\'{i}sica, Centro de Ci\^{e}ncias F\'{i}sicas e Matem\'{a}ticas, Universidade Federal de Santa Catarina, Campus Universit\'{a}rio Reitor Jo\~{a}o David Ferreira Lima, Florian\'{o}polis, Brazil, Zip Code: 88040-900}%
 
\author{D. Chinellato}
 \email{daviddc@g.unicamp.br}
 \affiliation{%
  Instituto de F\'isica Gleb Wataghin, Universidade Estadual de Campinas, R. S\'ergio Buarque de Holanda, 777, Campinas, Brazil, Zip Code: 13083-859
}

\author{A. V. Giannini}
 \email{giannini@ifi.unicamp.br}
 \affiliation{%
  Instituto de F\'isica Gleb Wataghin, Universidade Estadual de Campinas, R. S\'ergio Buarque de Holanda, 777, Campinas, Brazil, Zip Code: 13083-859
}

\author{M. N. Ferreira}
 \email{ansonar@uv.es}
 \affiliation{%
 Department of Theoretical Physics and IFIC,  University of Valencia and CSIC, E-46100, Valencia, Spain.
}%

\author{G. S. Denicol}
 \email{gsdenicol@id.uff.br}
\affiliation{%
  Instituto de F\'isica, Universidade Federal Fluminense,
  Av. Milton Tavares de Souza, Niter\'oi, Brazil, Zip Code: 24210-346,
}%

\author{M. Hippert}
 \email{hippert@illinois.edu}
 \affiliation{%
  Illinois Center for Advanced Studies of the Universe \& Department of Physics, 
University of Illinois at Urbana-Champaign, Urbana, IL 61801-3003, USA
}

\author{M. Luzum}
 \email{mluzum@usp.br}
\affiliation{
 Instituto de F\'isica, Universidade de S\~ao Paulo, R. do Mat\~ao, 1371, S\~ao Paulo, Brazil, Zip Code: 05508-090
}%

\author{J. Noronha}
 \email{jn0508@illinois.edu}
\affiliation{%
  Illinois Center for Advanced Studies of the Universe \& Department of Physics, 
University of Illinois at Urbana-Champaign, Urbana, IL 61801-3003, USA
}%

\author{J. Takahashi}
 \email{jun@ifi.unicamp.br}
 \affiliation{%
  Instituto de F\'isica Gleb Wataghin, Universidade Estadual de Campinas, R. S\'ergio Buarque de Holanda, 777, Campinas, Brazil, Zip Code: 13083-859
}

\collaboration{The ExTrEMe Collaboration}

\date{\today}

\begin{abstract}

We extend our previous investigation of the effects of pre-hydrodynamic evolution on final-state observables in heavy-ion collisions \cite{NunesdaSilva:2020bfs} to smaller systems. We use a state-of-the-art hybrid model for the numerical simulations with optimal parameters obtained from a previous Bayesian study. By studying p-Pb collisions, we find that the effects due to the assumption of a conformal evolution in the pre-hydrodynamical stage are even more important in small systems. We also show that this effect depends on the time duration of the pre-equilibrium stage, which is further enhanced in small systems. Finally, we show that the recent proposal of a free-streaming with subluminal velocity for the pre-equilibrium stage, thus effectively breaking conformal invariance, can alleviate the contamination of final state observables. Our study further reinforces the need for moving beyond conformal approaches in pre-equilibrium dynamics modeling, especially when extracting transport coefficients from hybrid models in the high-precision era of heavy-ion collisions.
\end{abstract}


\maketitle

\section{\label{sec:level1}Introduction}
Describing the behavior of strongly interacting matter under extreme conditions remains a challenge in high-energy physics. A transition from a confined hadronic phase to a deconfined phase, known as the quark-gluon plasma (QGP) \cite{Shuryak:1978ij}, at large densities and/or temperatures is predicted by quantum chromodynamics (QCD) \cite{Gross:1973id,Politzer:1973fx}, the fundamental theory of strong interactions. The strongly coupled nature of the problem and the sign problem that affects lattice QCD calculations \cite{deForcrand:2009zkb}, however, have hindered our ability to obtain an \textit{ab initio} theoretical determination of the complete phase diagram of QCD.

Aiming at shining a light on these questions from the experimental side, a vast program for producing and studying the QGP through relativistic heavy-ion collisions is being developed in facilities such as the Relativistic Heavy-Ion Collider (RHIC) and the Large Hadron Collider (LHC). By colliding pairs of Au and Pb nuclei, this program has produced mounting evidence that the matter formed in such collisions is well described by relativistic viscous hydrodynamics, and has  characterized the produced short-lived QGP as an expanding strongly-coupled fluid with a very small shear viscosity to entropy density ratio \cite{Heinz:2000bk,Arsene:2004fa,Back:2004je,Adcox:2004mh,Adams:2005dq,Muller:2012zq, ALICE:2012eyl, ATLAS:2012cix, CMS:2012qk, PHENIX:2013ktj, Heinz:2013th,Foka:2016vta, Loizides:2016tew}\footnote{In fact, the extracted values for this ratio are typically very close to the result $\eta/s=1/4\pi$, which is valid in a very large class of strongly-coupled gauge theories with holographic duals \cite{Kovtun:2004de}.}. 

In light of these findings, the QGP expansion in the aftermath of a collision is typically modeled within the framework of relativistic viscous hydrodynamics \cite{Israel:1976tn,Israel:1979wp}. In state-of-the-art simulations of relativistic heavy-ion collisions, hydrodynamics is one of the steps within a multimodel numerical chain that aims at modeling the several physical phenomena understood to take place throughout a collisional event. This approach is usually referred to as \textit{hybrid modeling} \cite{Petersen:2008dd} and has achieved success in reproducing experimental results for different colliding systems and different center of mass energies.

Some challenges, though, persist. One of them is to properly describe and model how the out-of-equilibrium matter produced immediately after the collision evolves into a state that can be described by hydrodynamics.

In practical simulations, the pre-hydrodynamic evolution has been modeled in various ways.  For example, with no transverse expansion of the system until some finite time when hydrodynamics begins \cite{Hirano:2002ds, Song:2007ux, Luzum:2008cw, Niemi:2015qia, Bozek:2009dw, Bernhard:2016tnd}, with the assumption of a type of zero-interaction free streaming of particles \cite{Liu:2015nwa, Bernhard:2019bmu, JETSCAPE:2020mzn, JETSCAPE:2020shq}, or via classical Yang-Mills theory \cite{Schenke:2012wb}. In these cases, the question of thermalization is side-stepped, as thermalization is never reached, but is either imposed by hand at the switching to a hydrodynamic description or taken care of within the hydrodynamic evolution, where the system evolves from a far-from-equilibrium state into one that is closer to equilibrium.  In other simulations (see, for example,  \cite{vanderSchee:2013pia}), an evolution toward equilibrium is obtained via a strongly-coupled description in terms of the gauge-gravity duality \cite{Maldacena:1997re}, or in a weak-coupling approach via effective kinetic theory \cite{Kurkela:2018vqr, Kurkela:2018wud}.  It is worth noting that, for simplicity, the vast majority of these approaches assume an approximate conformal symmetry during this pre-hydrodynamic evolution. Furthermore, the modeling of the pre-equilibrium phase is also very important when determining the validity of the hydrodynamic description using nonlinear causality constraints \cite{Plumberg:2021bme}.  

In a previous work \cite{NunesdaSilva:2020bfs}, we have studied different pre-equilibrium models and their effects on several final state observables. We have pointed out that the investigated models all added extra momentum to the system, in comparison to a base scenario in which the period of pre-equilibrium evolution was replaced by hydrodynamic evolution itself.

We have argued that this non-negligible extra momentum is not due to interactions among the constituents of the early-time system, but instead is a consequence of an artificial large pressure that is present at the switch to hydrodynamics due to the widespread assumption of conformal invariance in pre-equilibrium models. This artifact ultimately affects the extraction of transport coefficients from hybrid models and should be taken as a caveat in Bayesian studies.

Another pressing issue in the field regards the validity of the hydrodynamical description for smaller systems. Different experimental collaborations have presented data exhibiting strong signatures of  flow in p-Pb and d-Au collisions \cite{ALICE:2012eyl, ATLAS:2012cix, CMS:2012qk, PHENIX:2013ktj}. As reviewed in \cite{Loizides:2016tew}, the hydrodynamical picture of the QGP is capable of describing not only flow observables but a wide collection of experimental data from such small systems. From the theoretical point of view, different arguments \cite{Habich:2015rtj, Spalinski:2016fnj}, mostly based on the strongly coupled nature of the QGP, support the validity of the hydrodynamical picture in the description of small collision systems. 

In this work, we extend our previous study on the effects of pre-equilibrium dynamics on final-state observables to include smaller systems. We study simulated collision events for both p-Pb and Pb-Pb, utilizing a numerical setup inspired by the one used in the Bayesian study by Bernhard \textit{et al.} \cite{Moreland:2018gsh} to obtain maximum a posteriori values for parameters of the initial conditions and hydrodynamics viscosities. We show that, as  expected, the effect due to the unphysically large initial pressure, originating from the assumption of conformal symmetry, is enhanced as we move to smaller systems, and appropriately removing it becomes even more important in studies of such systems. We also show that this effect is dependent on the duration of the pre-conformal stage. Finally, we study whether using a subluminal free-streaming model for the pre-equilibrium stage, which effectively breaks conformal symmetry, can alleviate the effects related to the aforementioned artifact.

In the next section, we review the non-physical enhancement of the bulk pressure at the transition to hydrodynamics due to the widespread assumption of conformal evolution in pre-equilibrium dynamics models; in Section~\ref{sec:hybrid} we present our numerical setup. We assume zero chemical potential in this work. Our results are presented in Section~\ref{sec:results}. This is followed by our conclusions. We use a mostly minus metric signature and natural units $\hbar=c=k_B=1$.
\section{(Non)Conformal pre-equilibrium and the switch to hydrodynamics}

We have argued in previous work \cite{NunesdaSilva:2020bfs} that the conformal invariance assumed by most initial conditions and pre-equilibrium models \cite{Schenke:2012wb, Schenke:2012fw, Bernhard:2018hnz,Bernhard:2019bmu, Romatschke:2015gxa, Weller:2017tsr, JETSCAPE:2020mzn} leads to an artificially large bulk pressure at the time one switches to the hydrodynamic description. This extra pressure significantly modifies the transverse momentum spectra of final state particles, contaminating integrated final state observables and the extraction of transport coefficients from Bayesian analysis. We now review this argument and expand on it, in light of the recent proposition of a subluminal free-streaming period as a pre-equilibrium model \cite{Nijs:2020ors, Nijs:2020roc}.

We remind the reader that, at the end of the pre-equilibrium evolution at time $\tau_{fs}$, the switch to hydrodynamics is performed 
by ensuring continuity of the entire stress-energy tensor $T^{\mu\nu}$.
The fluid-dynamical variables are defined via Landau matching \cite{LandauLifshitzFluids}, which determines the fluid's flow velocity $u^\mu$ through the eigenvalue problem
\begin{equation}
    T^\mu_\nu u^\nu \equiv \epsilon u^\mu. \label{eigen} 
\end{equation}
The single time-like eigenvector, $u^\mu$, is selected and normalized to $u^\mu u_\mu = 1$ and its corresponding eigenvalue $\epsilon$ is identified with the energy density. The energy-momentum tensor can then be decomposed as
\begin{equation}
    T^{\mu\nu} = \epsilon u^\mu u^\nu - [ p(\epsilon) + \Pi ]\Delta^{\mu\nu} + \pi^{\mu\nu} \,, 
    \label{eqn:LandauDecomp}    
\end{equation} 
where $p(\epsilon)$ is the thermodynamic pressure, related to the energy density through an equation of state (EOS), $\Pi$ is the bulk pressure, and $\pi^{\mu\nu}$ the shear stress tensor, 
\begin{equation}
    \pi^{\mu\nu} = \Delta^{\mu\nu}_{\alpha\beta}T^{\alpha\beta} \,,    
\end{equation} 
which is traceless. The projector 
\begin{equation}
    \Delta^{\mu\nu} = g^{\mu\nu} - u^\mu u^\nu \,,    
\end{equation}
projects out terms transverse to $u^\mu$, and we define
\begin{equation}
    \Delta^{\mu\nu}_{\alpha\beta} = \frac{1}{2}\left( \Delta^\mu_\alpha \Delta^\nu_\beta + \Delta^\mu_\beta \Delta^\nu_\alpha \right) - \frac{1}{3}\Delta^{\mu\nu}\Delta_{\alpha\beta} \,.    
\end{equation} 
Taking the trace of \eqref{eqn:LandauDecomp}, the total pressure may be obtained as
\begin{equation}
    p(\epsilon) + \Pi = \frac{ \epsilon - T^\mu_\mu }{3}\,, 
    \label{eqn:PifromT_alt}    
\end{equation}
In the case of a conformal pre-equilibrium model, $T^\mu_\mu = 0$ (regardless of the history of the system or proximity to equilibrium), and
\begin{equation}
     p(\epsilon) + \Pi= \frac{\epsilon}{3}.
\label{eqn:TmunuDecompositionConformal}
\end{equation}
That is, at the switching time, the conformal pressure $\epsilon/3$ must be matched to the total QCD pressure. Since QCD is not conformal, the difference between the thermodynamic and conformal pressures must be accommodated by an artificial bulk pressure. 
\begin{figure}[!ht]
  \includegraphics[width=.5\linewidth]{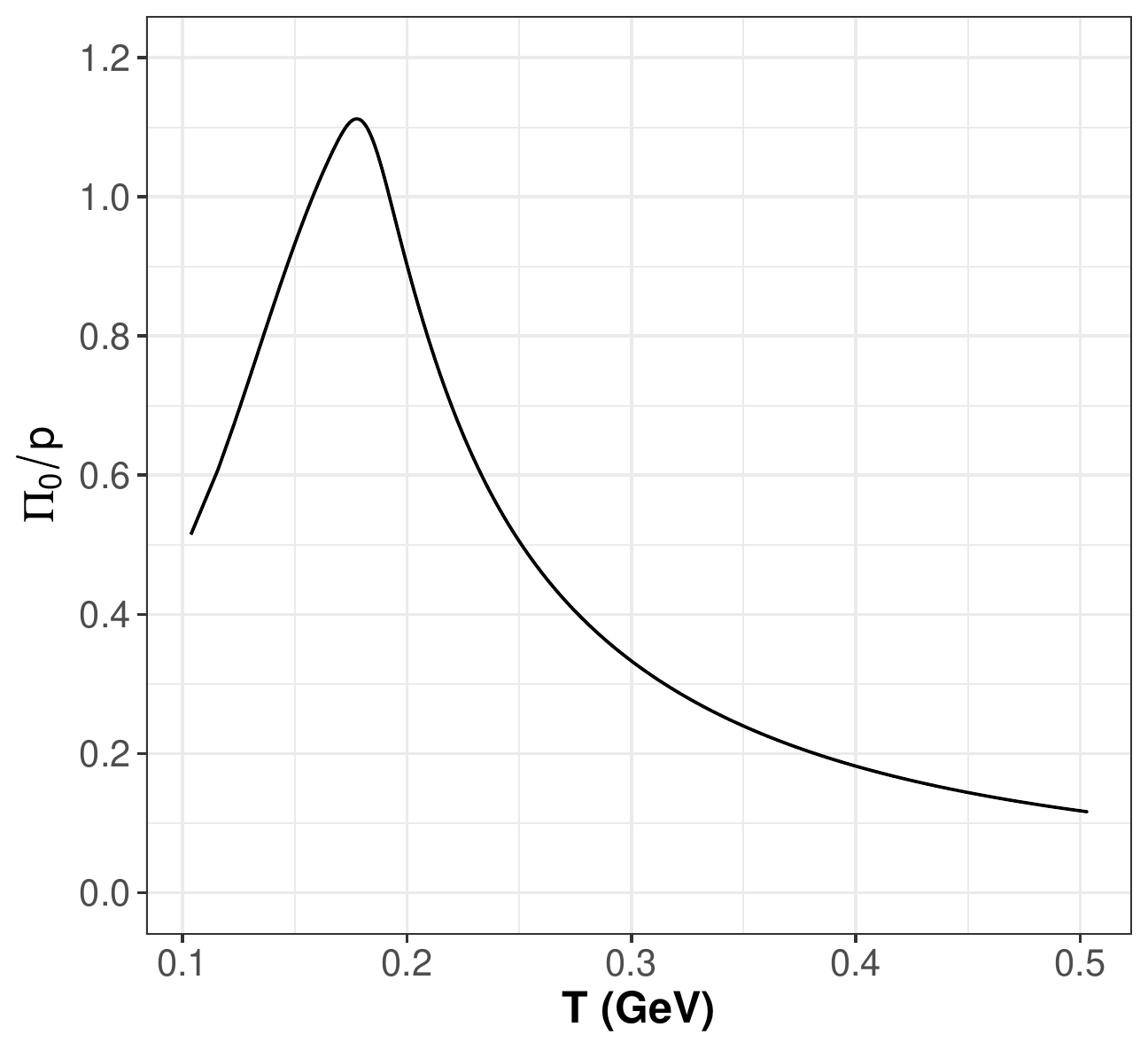}
\caption{Ratio between the initial bulk pressure $\Pi_0$ and QCD thermodynamic pressure $p(e)$ calculated from the HotQCD equation of state \cite{HotQCD:2014kol} matched to the particle content of UrQMD while requiring that $T^{\mu}_\mu=0$ at the hydrodynamization time. The result is plotted as a function of the equivalent QCD temperature $T$. See Eq.~\eqref{eqn:TmunuDecompositionConformal}.}
  \label{fig:bulk-over-p}
\end{figure}
In Figure~\ref{fig:bulk-over-p}, we present the ratio between the initial bulk pressure $\Pi_0$ and the QCD thermodynamic pressure as a function of temperature, for the range of temperatures typically probed in hybrid simulations of relativistic heavy-ion collisions. It can be seen that this ratio reaches values of order unity, meaning that the nonphysically enhanced initial bulk viscous pressure can have the same order of magnitude as the actual thermodynamic pressure for some cells at the beginning of hydrodynamics. 

Recently, it was proposed by Nijs \textit{et al.} \cite{Nijs:2020ors, Nijs:2020roc} to describe the pre-equilibrium dynamics using a free-streaming model with a variable free-streaming velocity that can take subluminal values, effectively breaking conformal invariance. Choosing Milne coordinates with the metric being given by
\begin{equation}
    g_{\mu\nu} = {\rm diag}(1, -1, -1, -\tau^2 ) \,,
\end{equation}
and assuming a boost invariant system, the energy-momentum tensor of the QGP along the transverse plane (with respect to the beam axis) can be written as
\begin{equation}
    T^{\mu\nu}(x,y) = \frac{1}{2\pi\tau_{fs}}\int_0^{2\pi} d\phi \, \hat{p}^\mu \hat{p}^\nu \, \mathcal{T}( x - \vfs\tfs\cos\phi, y - \vfs\tfs\sin\phi ) \,, 
    \label{eqn:Tmunu}
\end{equation}
where $\tau_{fs}$ is the total free-streaming time of the system
and 
\begin{equation}
    \hat{p}^\mu \hat{p}^\nu = 
    \begin{pmatrix}
    1 & \vfs\cos \phi & \vfs\sin\phi & 0 \\
    \vfs\cos \phi & \vfs^2\cos^2\phi & \vfs^2\cos\phi \sin \phi & 0\\
    \vfs \sin \phi & \vfs^2 \cos\phi \sin \phi & \vfs^2 \sin^2\phi & 0 \\
    0 & 0 & 0 & 0
    \end{pmatrix} \,,
\end{equation}
such that $T^{\mu\nu}$ has vanishing $\eta$ components.
${\cal T}(x,y)$ is related to the probability density function of the Boltzmann equation after integrating out the longitudinal degrees of freedom~\cite{Liu:2015nwa} and can be identified with an initial density profile, generated, for the example, using \trento~\cite{Moreland:2014oya}.

Note that the trace $T^\mu_\mu(x,y)$ is given by
\begin{equation} 
    T^\mu_{\;\;\mu}(x,y) = \frac{(1 - \vfs^2)}{2\pi\tfs}\int_0^{2\pi} d\phi \, \mathcal{T}( x - \vfs\tfs\cos\phi, y - \vfs\tfs\sin\phi ) \neq 0\,,
    \label{eqn:trace_anom}
\end{equation}
indeed breaking conformal invariance. Note, also, that for $\vfs = 1$, the conformal limit is recovered, with $T^\mu_\mu = 0$.

\begin{figure}[!ht]
    \includegraphics[width=.475\linewidth]{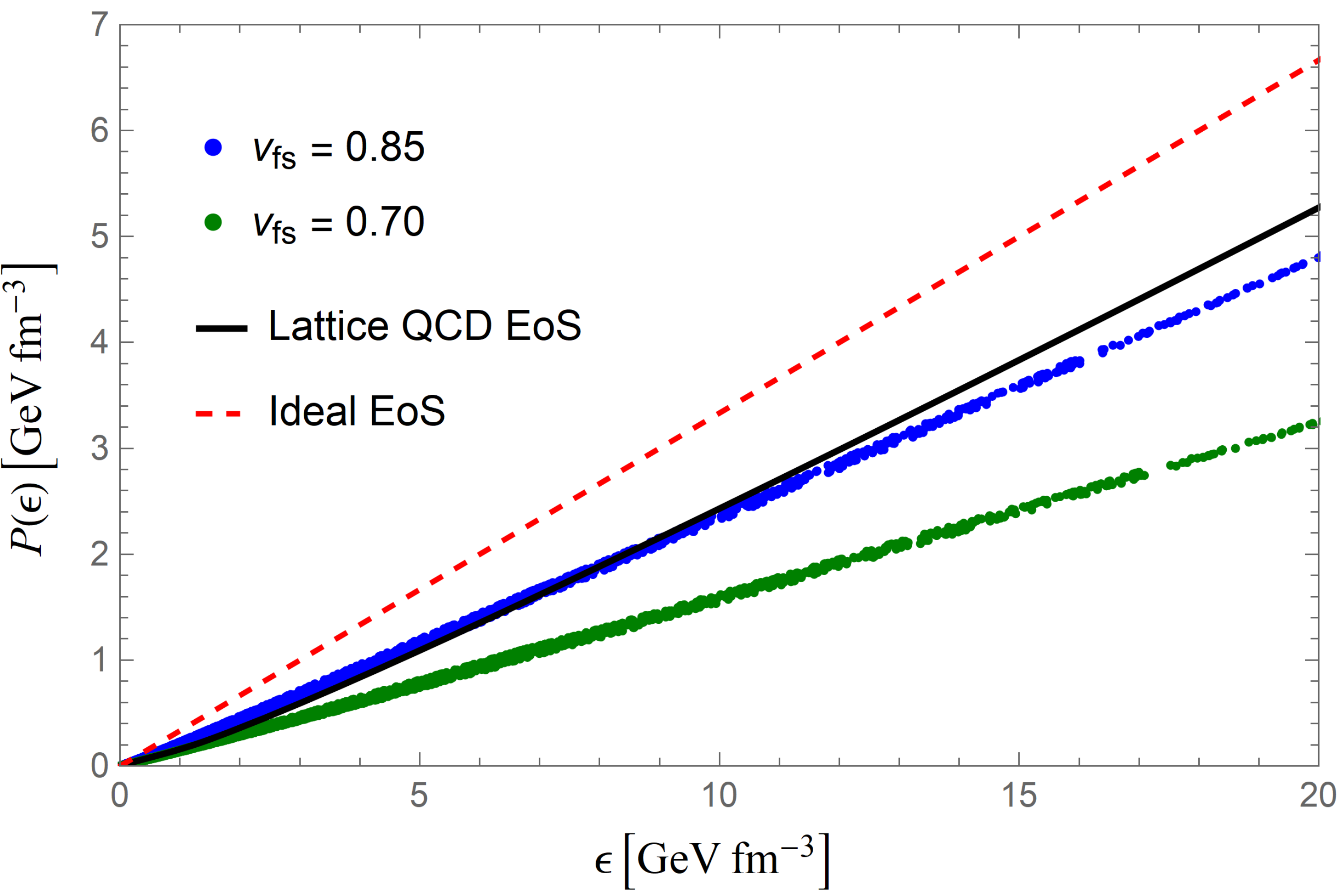}
    \caption{Total pressure as a function of energy, obtained by numerically solving Eq.~\eqref{eqn:pressureEqn}. Results are presented for different free-streaming velocities.}
    \label{fig:P_FS}
\end{figure}

Using \eqref{eqn:trace_anom} for the trace anomaly into \eqref{eqn:PifromT_alt} we obtain the explicit expression 
\begin{equation}
    p(\epsilon) + \Pi = \frac{ \epsilon }{3} - \frac{(1 - \vfs^2)}{6\pi\tfs}\int_0^{2\pi} d\phi \, \mathcal{T}( x - \vfs\tfs\cos\phi, y - \vfs\tfs\sin\phi ) \,.
    \label{eqn:pressureEqn}
\end{equation}
Equation~\eqref{eqn:pressureEqn} can be solved numerically to yield the total pressure for non-conformal free-streaming. Some results are presented in Figure~\ref{fig:P_FS} for different free-streaming velocities. 

We note that, for $v=0.85\,c$, the free-streaming pressure approximates the QCD equation of state for a considerable range of energies. According to the arguments presented in \cite{NunesdaSilva:2020bfs} and reviewed above, a subluminal free-streaming pre-equilibrium model should result in a smaller artificial bulk pressure, alleviating the contamination of final state observables. 

To explore that possibility, we have incorporated such a free streaming model into a simulation chain inspired by the one used in \cite{Nijs:2020ors, Nijs:2020roc}, as we will describe in the next section. We will also use the resulting model to extend our previous analysis to systems of smaller size and to probe how the effects of the conformal approximation change with the duration of the free-streaming period.

\section{\label{sec:hybrid} Numerical Model for Hybrid Description of Heavy-Ion Collisions}
As mentioned above, hybrid models are the state-of-the-art tool for simulating relativistic heavy-ion collisions. They are most typically comprised of the following components:
\begin{itemize}
    \item an initial condition generator, which models the initial hard scattering between nuclei \cite{Bass:1998ca, Bleicher:1999xi, Miller:2007ri,Loizides:2014vua,Zhang:1999bd, Lin:2004en, Werner:2005jf, PhysRevC.65.054902, Schenke:2012wb,Schenke:2012fw,Moreland:2014oya, Weil:2016zrk};
    \item a pre-equilibrium stage, which models matter approach to hydrodynamical behavior \cite{Broniowski:2008qk,Schenke:2012wb,Schenke:2012fw,Liu:2015nwa,Kurkela:2018vqr, Kurkela:2018wud};
    \item a relativistic viscous hydrodynamics code, for modeling the hydrodynamical evolution of the QGP \cite{Baier:2006gy,Song:2007ux, Noronha-Hostler:2013gga, Noronha-Hostler:2014dqa, Shen:2014vra, Schenke:2010nt, Schenke:2011bn,Paquet:2015lta};
    \item a Cooper-Frye sampler, for converting fluid elements into particles \cite{PhysRevD.10.186,Pratt:2010jt,Huovinen:2012is,Pratt:2014vja};
    \item a hadronic cascade model, for modeling the final evolution of hadron resonances up to the detectors \cite{Bass:1998ca, Bleicher:1999xi,Chojnacki:2011hb,Petersen:2018jag,Mazeliauskas:2018irt}.
\end{itemize}

In this work, simulations of collision events are generated using this type of hybrid model. In practice, 2D initial entropy profiles are generated with version 2.0 of \trentonosp\cite{Moreland:2014oya}. They are then free-streamed, with velocity $v_{fs}$, up to a time $\tau_{fs}$, at which the hydrodynamical evolution is performed with the MUSIC code \cite{Schenke:2010nt, Schenke:2011bn,Ryu:2015vwa,Paquet:2015lta}. MUSIC performs a boost invariant $2D+1$ viscous evolution of the system, which ends when the systems reaches a temperature of $T_\text{switch} = 151$ MeV. Following \cite{Moreland:2018gsh}, we have parametrized the shear viscosity to entropy density ratio as,
\begin{equation}
    (\eta/s)(T) = (\eta/s)_{\text{min}} + (\eta/s)_{\text{slope}} \cdot (T-T_c) \cdot (T/T_c)^{(\eta/s)_{\text{crv}}}.
\end{equation}
In the expression above, $(\eta/s)_{\text{min}}$ represents the minimum value at $T_c$, $(\eta/s)_{\text{slope}}$ the slope above $T_c$, and $(\eta/s)_{\text{crv}}$ is a curvature parameter. The bulk viscosity to entropy density ratio is parametrized by
\begin{equation}
    (\zeta/s)(T) = \frac{(\zeta/s)_{\text{max}}}{1 + \left( \frac{T - (\zeta/s)_{T_0}}{(\zeta/s)_{\text{width}}}\right)^2},
\end{equation}
which has the form of a Cauchy distribution, with a symmetric peak with maximum value $(\zeta/s)_{\text{max}}$, width $(\zeta/s)_{\text{width}}$, and center $(\zeta/s)_{T_0}$. 

The equation of state used in the hydrodynamical stage is obtained by matching the HotQCD lattice equation of state \cite{HotQCD:2014kol} to a hadron resonance gas composed of the same particle species present in UrQMD \cite{Bass:1998ca, Bleicher:1999xi}, using the EOS-maker software \cite{eosMaker}.

The final hydrodynamic hypersurface (at constant temperature) is sampled, via the Cooper-Frye formalism with viscous corrections using the \textsc{frzout} code \cite{Moreland:2018gsh}. Each hypersurface is repeatedly sampled until at least $5\times 10^5$ particles are acquired. The resulting set of particles is used as input into the afterburner wrapper for the UrQMD hadronic cascade model \cite{Bass:1998ca, Bleicher:1999xi}. The final set of stable particles is then used for the calculation of observables.

In order to investigate pre-equilibrium dynamics effects in large and small systems, we have simulated minimum-bias events for both Pb-Pb and p-Pb collisions at center of mass energy of $5.02$ TeV. We have simulated sets of events for two different values of $\tau_{fs}$, namely $\tau_{fs} = 0.37$ fm/$c$ and $\tau_{fs} = 1.2 $ fm/$c$. To investigate the effectively conformally broken scenario, for both values of $\tau_{fs}$ we have considered two different values of $v_{fs}$: $v_{fs} = c$ and $v_{fs} = 0.85\,c$.

The $v_{fs} = c$ and $\tau_{fs} = 0.37$ fm/$c$ scenario matches the maximum a posteriori parameters obtained from the Bayesian analysis performed in \cite{Moreland:2018gsh}. All of the other parameters for the \trentonosp initial conditions and hydrodynamics transport coefficients were extracted from this reference for this base scenario. For the other scenarios, we have adjusted the overall \trentonosp normalization factor, to obtain agreement across the scenarios in charged-particle multiplicity at the most central class.

The choice of $v_{fs} =  0.85$ fm/$c$ as the second free-streaming velocity is motivated by the works of \cite{Nijs:2020ors, Nijs:2020roc} and by the results presented in Fig.~\ref{fig:P_FS}. While we do not expect the observables extracted from simulations with this velocity to match experimental data, we shall see that this remains a robust choice for probing the consequences of effectively breaking conformal invariance in the pre-equilibrium stage of the hybrid description.

The parameters utilized in this work are summarised in Table~\ref{tab:parameters} and the normalization factors for the different scenarios are summarised in Table~\ref{tab:normalizations}.
\begin{center}
    \begin{table}
        \begin{tabular}{ |p{2cm}|p{2cm}||p{2cm}|p{2cm}|  }
            \hline
            \multicolumn{2}{|c|}{Initial Condition} & \multicolumn{2}{|c|}{QGP Properties} \\
            \hline
            &  & $(\eta/s)_\text{min}$ & 0.11\\
            $v$ & 0.43 & $(\eta/s)_\text{slope}$ & 1.6 (1/GeV) \\
            $n_c$ & 6 & $(\eta/s)_\text{cvr}$ & -0.29\\
            $p$ & 0.0 & $(\zeta/s))_\text{max}$ & 0.032\\
            $k$ & 0.19 & $(\zeta/s))_\text{width}$ & 0.024 GeV\\
            $w$ & 0.92 fm & $(\zeta/s))_{T_0}$ & 175 MeV\\
            $d_\text{min}$ & 0.81 fm  & $T_\text{swicth}$ & 151 MeV\\
            \hline
        \end{tabular}
        \caption{\label{tab:parameters} Parameters used for initial conditions generation and transport coefficients for hydrodynamical evolution.}
    \end{table}
\end{center}
\begin{table}[]
    \begin{tabular}{|p{3cm}|p{2cm}|p{2cm}|}
        \hline
        & $v_{fs} = 0.85 c$ & $v_{fs} = c$ \\ \hline
        $\tau_{fs} = 0.37$ fm/$c$ & 20.0               & 20.0          \\
        $\tau_{fs} = 1.2$ fm/$c$  & 22.0               & 26.8          \\ \hline
    \end{tabular}
    \caption{\label{tab:normalizations} Overall normalization factors used in \trentonosp for initial condition generation for the different scenarios under study.}
\end{table}

\section{\label{sec:results} Results and Discussion}
\begin{figure}[!ht]
  \includegraphics[width=.475\linewidth]{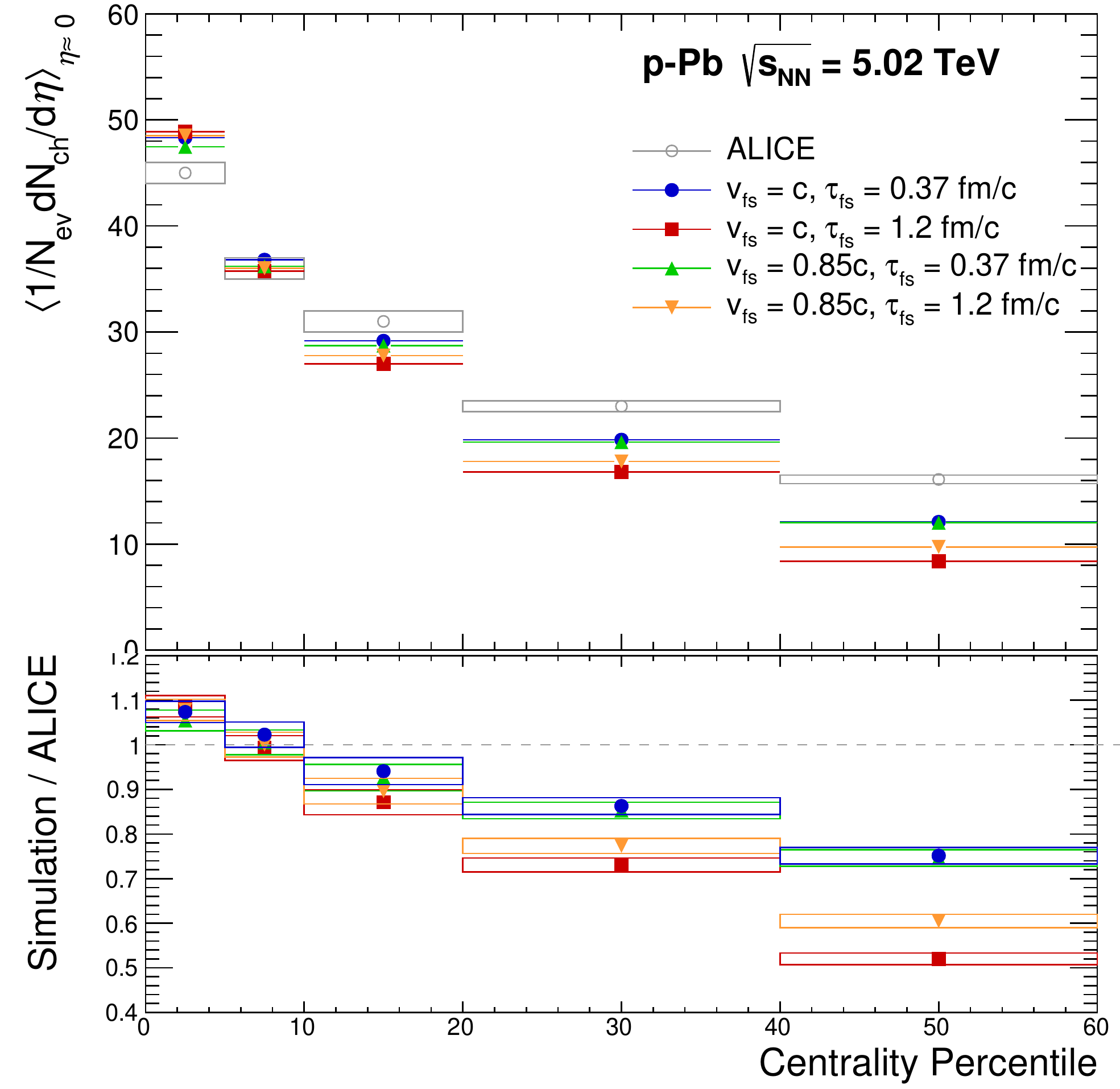}
  \includegraphics[width=.475\linewidth]{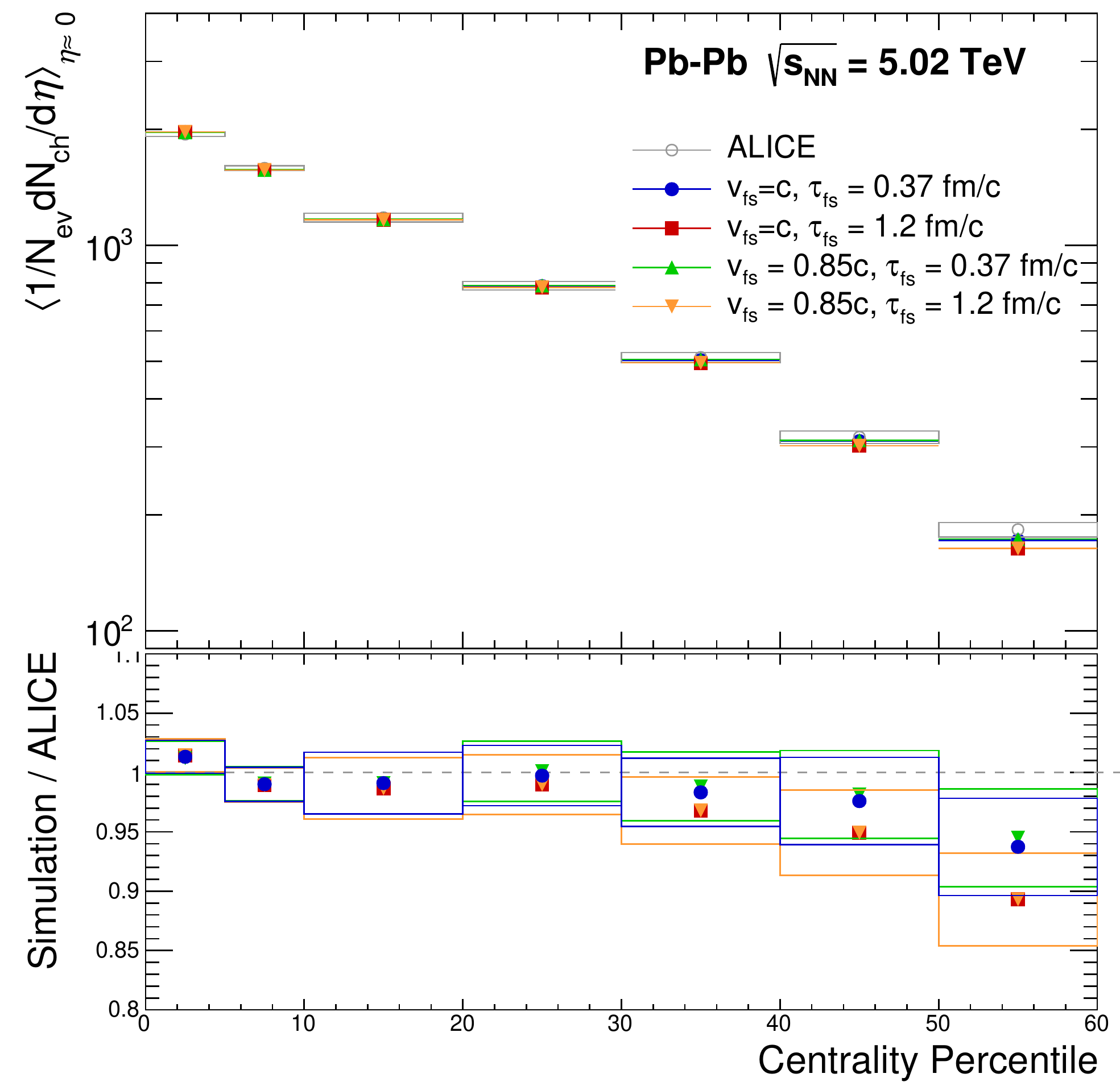}
\caption{Final multiplicity of charged particles as a function of centrality for all of the pre-hydrodynamical scenarios under consideration (top panel). Results are compared to experimental data from the ALICE Collaboration \cite{ALICE:2013wgn, ALICE:2015juo}. We also show the ratio between the experimental results and results from our  simulations (bottom panel).}
  \label{fig:mult}
\end{figure}

First, we present the resulting charged-particle multiplicity as a function of centrality class for each of the simulated scenarios in Fig.~\ref{fig:mult}, where they are compared to experimental results from the ALICE collaboration~\cite{ALICE:2013wgn,ALICE:2015juo}. We obtain comparable results as those obtained by the original Bayesian study \cite{Moreland:2018gsh} for the scenarios with $v_{fs} = c$ and $\tau_{fs} = 0.37$ fm/$c$, for both p-Pb and Pb-Pb collisions. We also note that the multiplicity dependence on the choice of $\tau_{fs}$ and $v_{fs}$ is considerably larger for the smaller system. The dependence on $\tau_{fs}$ is also stronger than the one on $v_{fs}$. 

\begin{figure}[!ht]
    \includegraphics[width=.425\linewidth]{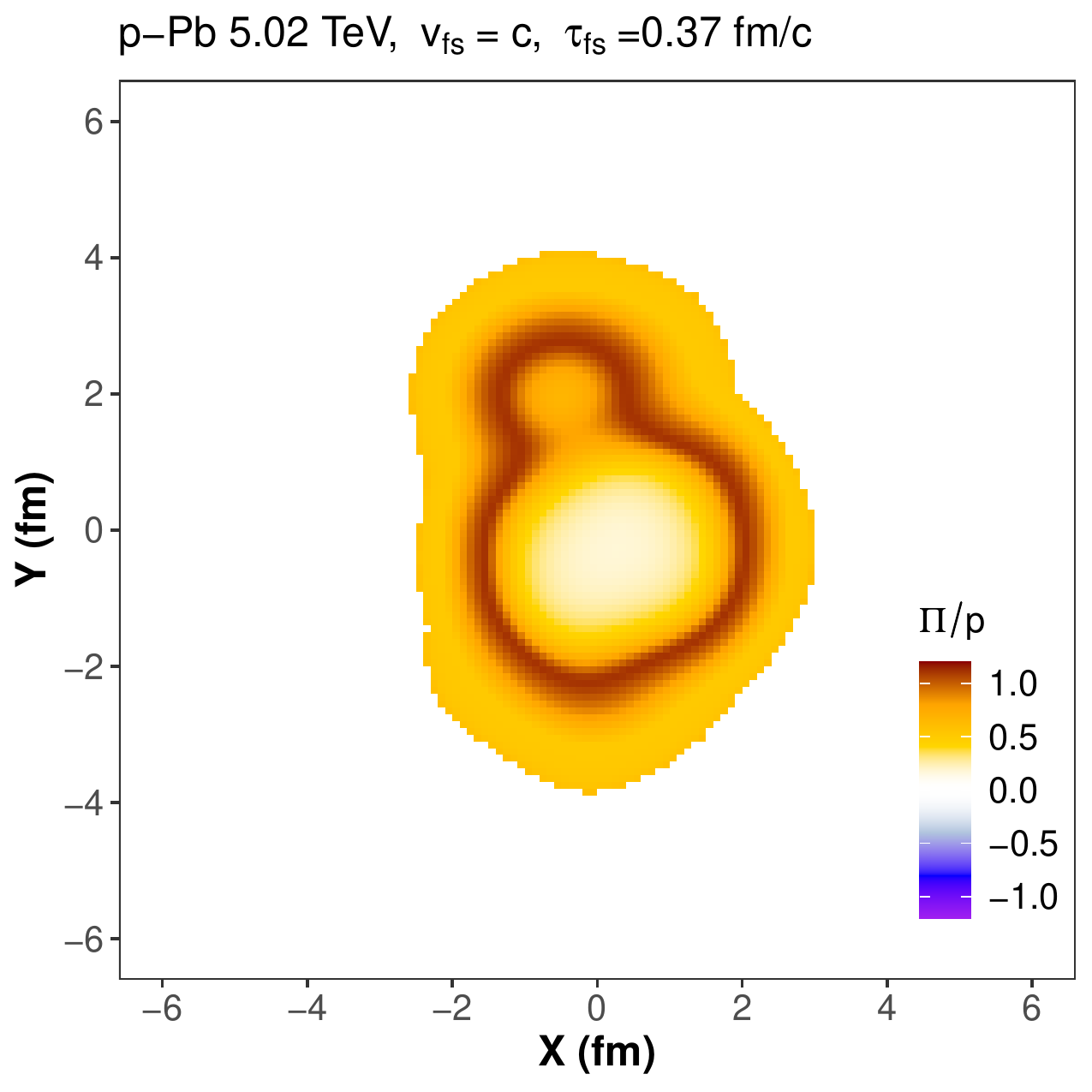}
    \includegraphics[width=.425\linewidth]{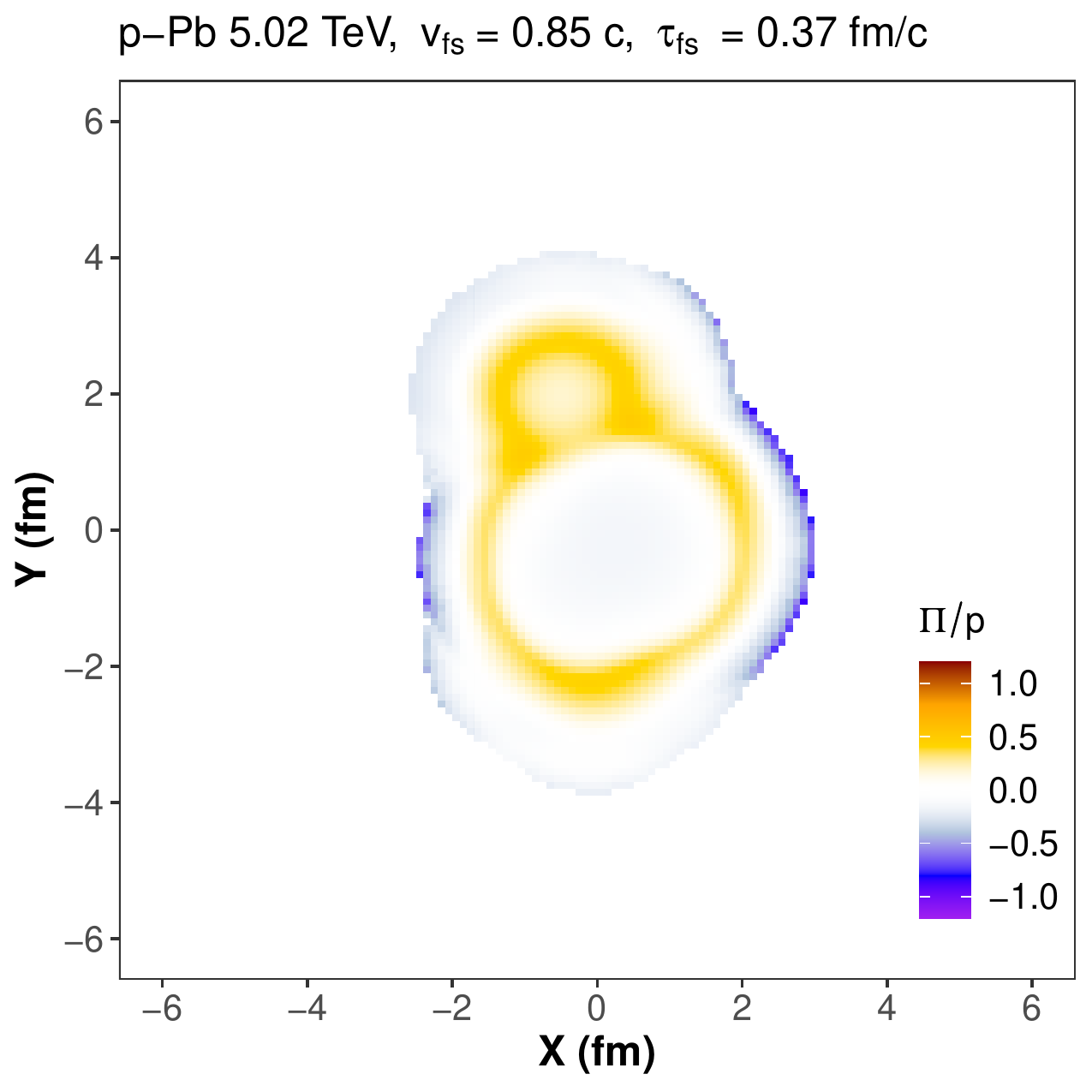}
    \includegraphics[width=.425\linewidth]{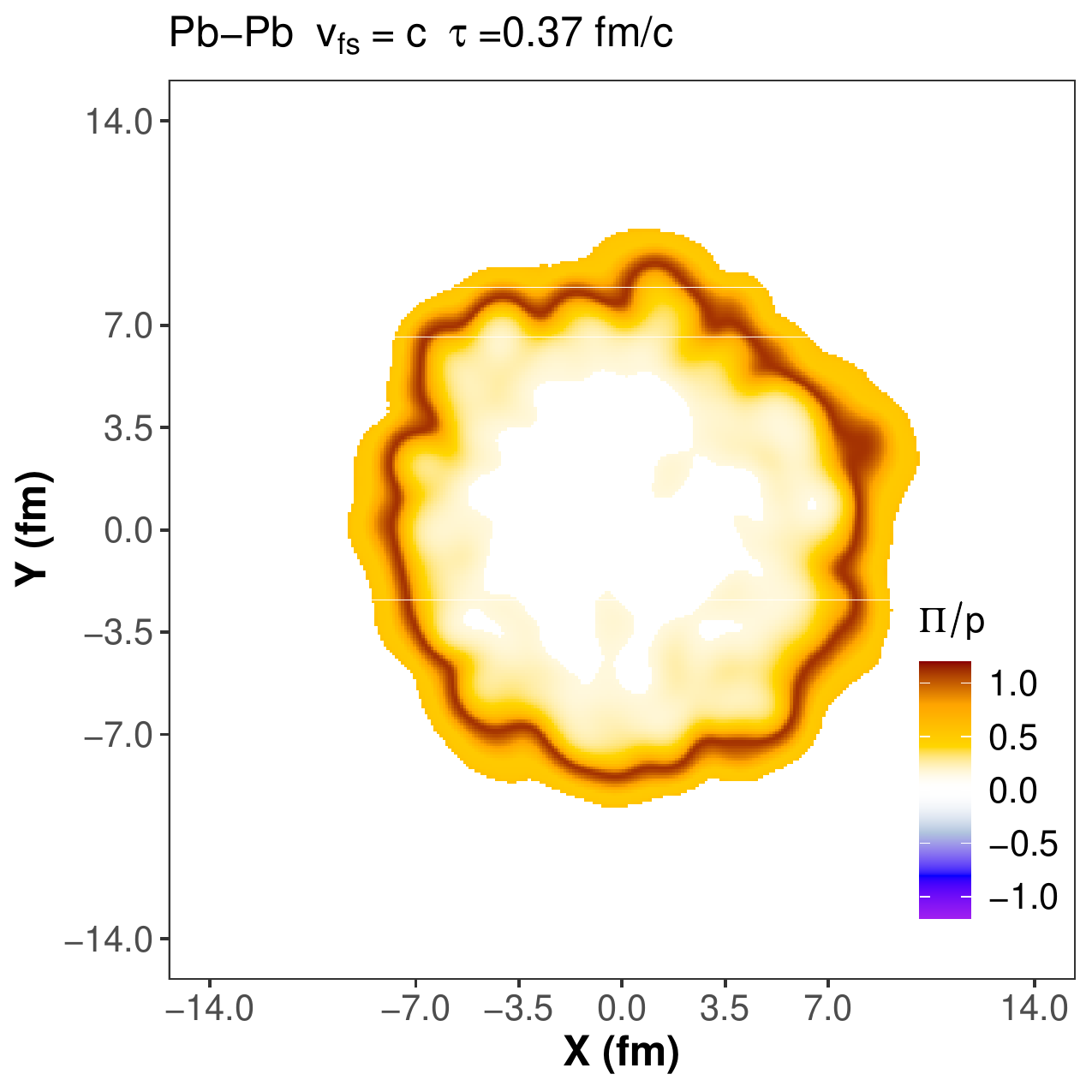}
    \includegraphics[width=.425\linewidth]{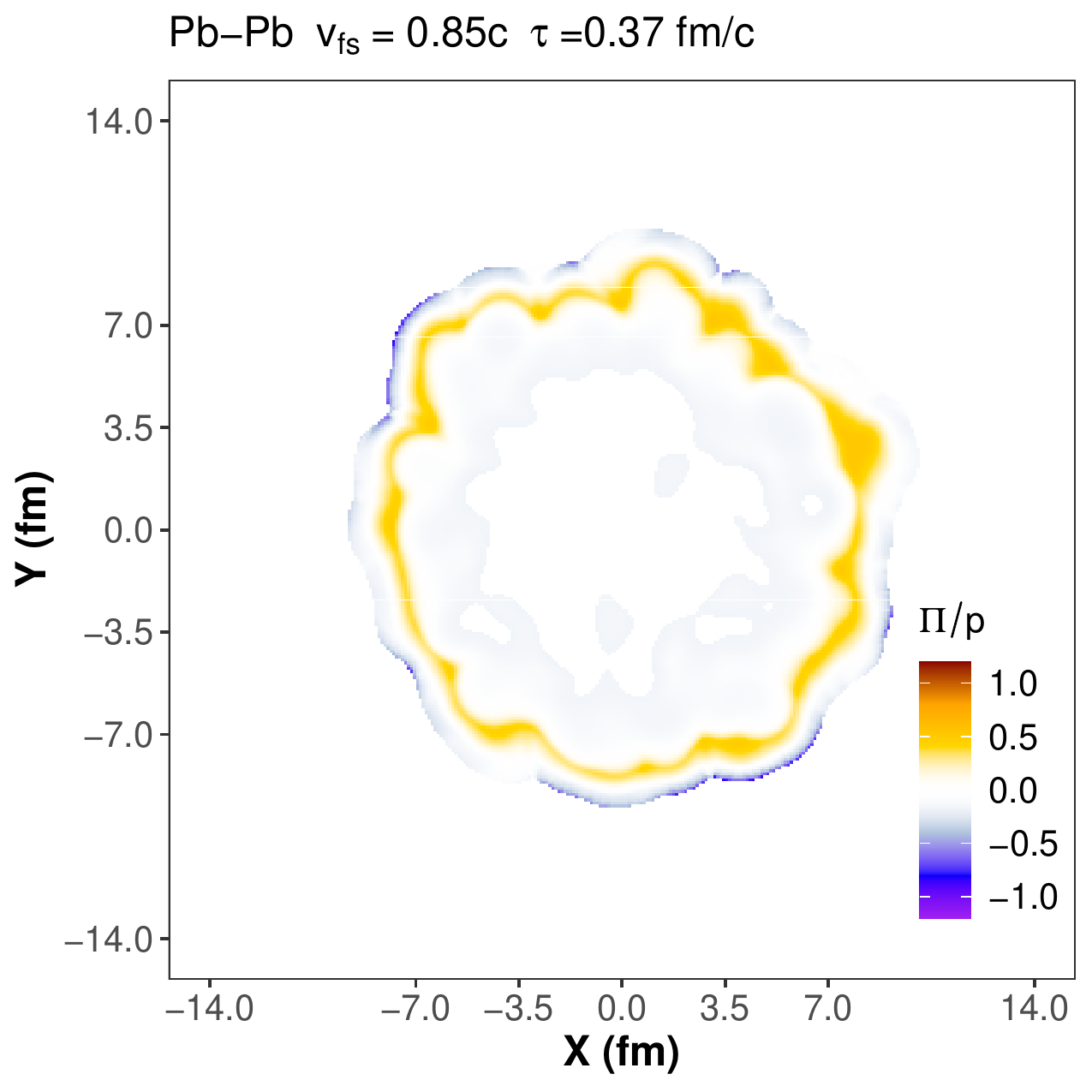}
    \caption{Ratio $\Pi/p$ between the bulk pressure $\Pi$ and the thermodynamic pressure $p$ across sample p-Pb and Pb-Pb events.}
    \label{fig:bulkP}
\end{figure}

In Figure~\ref{fig:bulkP} we compare the ratio $\Pi/p$ in sample central events at the beginning of the hydrodynamic evolution ($\tau=\tau_{fs}$), for $v_{fs} = c$ and $v_{fs} = 0.85\,c$ and for both system sizes. Note that for $v_{fs}=c$ a significantly larger fraction of the p-Pb event has $\Pi/p \sim \mathcal{O}(1)$, when compared to the Pb-Pb event, suggesting that the effects of the artificial large bulk pressure on the final state observables should be larger on the smaller system.  We also remark that for $v_{fs}=0.85\,c$ these ratios seem to be greatly diminished in most cells, pointing to a reduction of the above-mentioned effect.

To investigate and quantify the effect of the unphysical enhancement of initial bulk viscous pressure in small systems and compare it to the previously observed behavior in large systems, we have calculated several final state observables across the scenarios under investigation. A key observable to be investigated, as noted in \cite{NunesdaSilva:2020bfs}, is the transverse momentum spectra for both p-Pb and Pb-Pb simulated events.
\begin{figure}[!ht]
  \includegraphics[width=.475\linewidth]{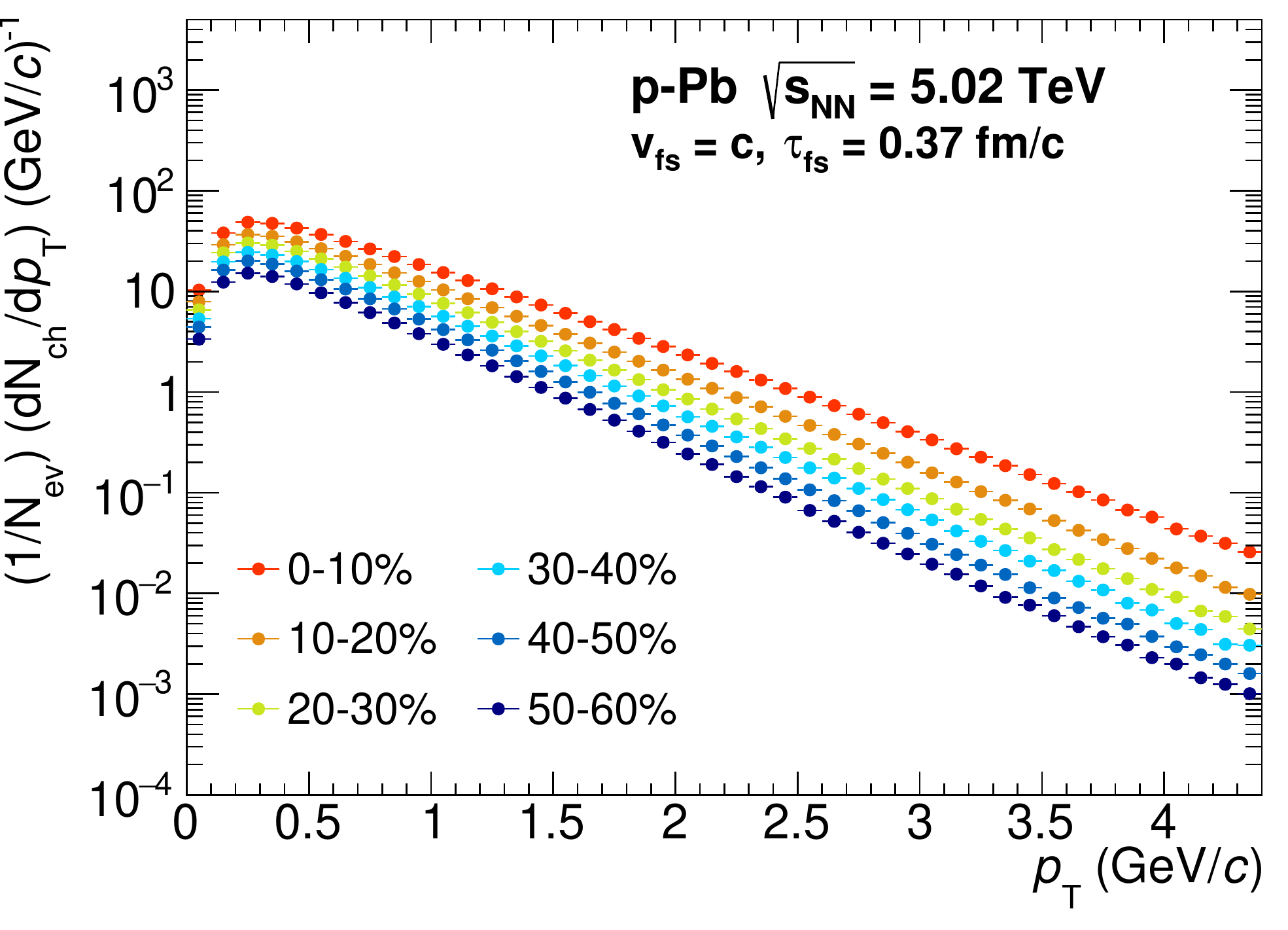}
  \includegraphics[width=.475\linewidth]{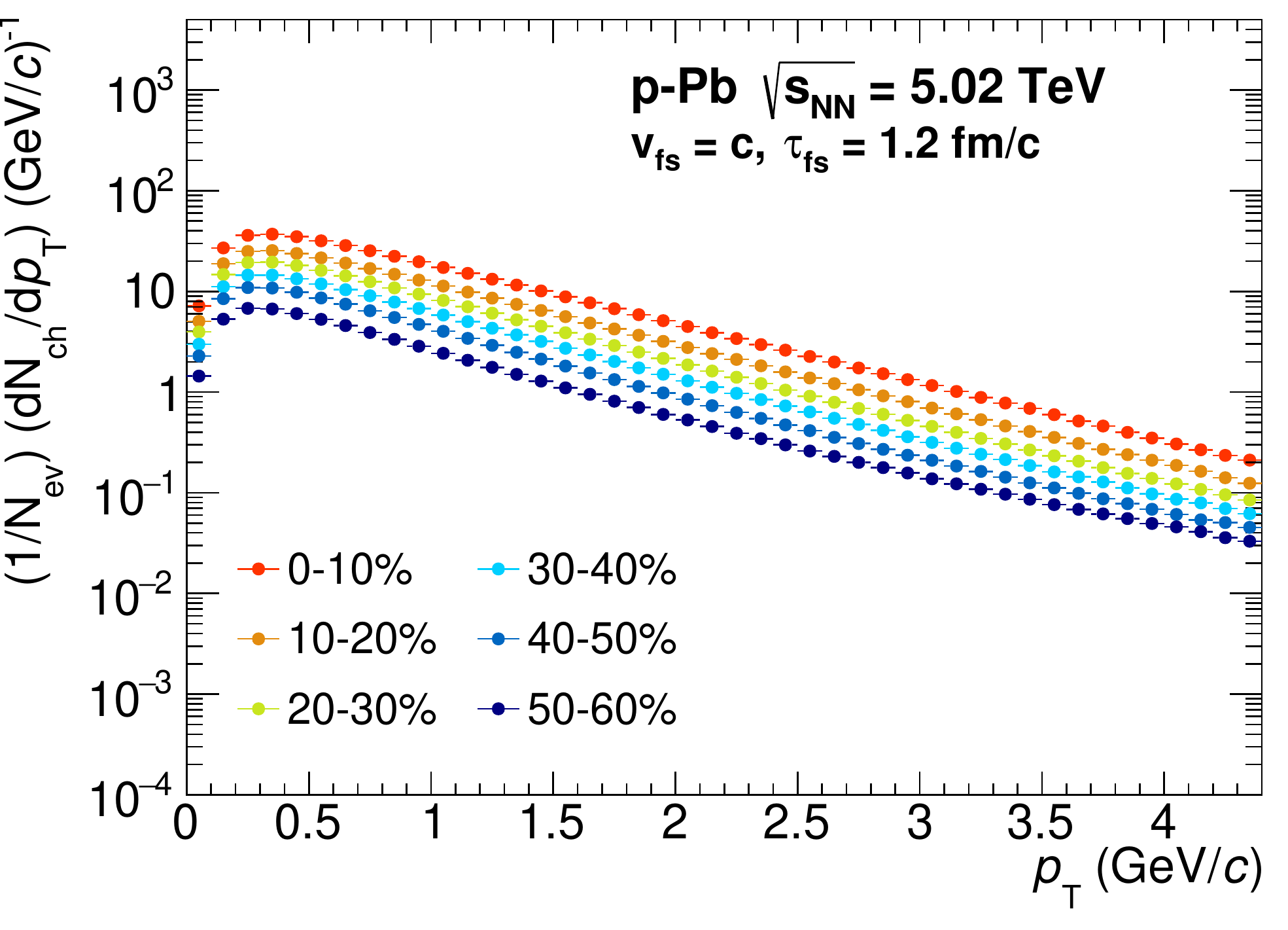}
  \includegraphics[width=.475\linewidth]{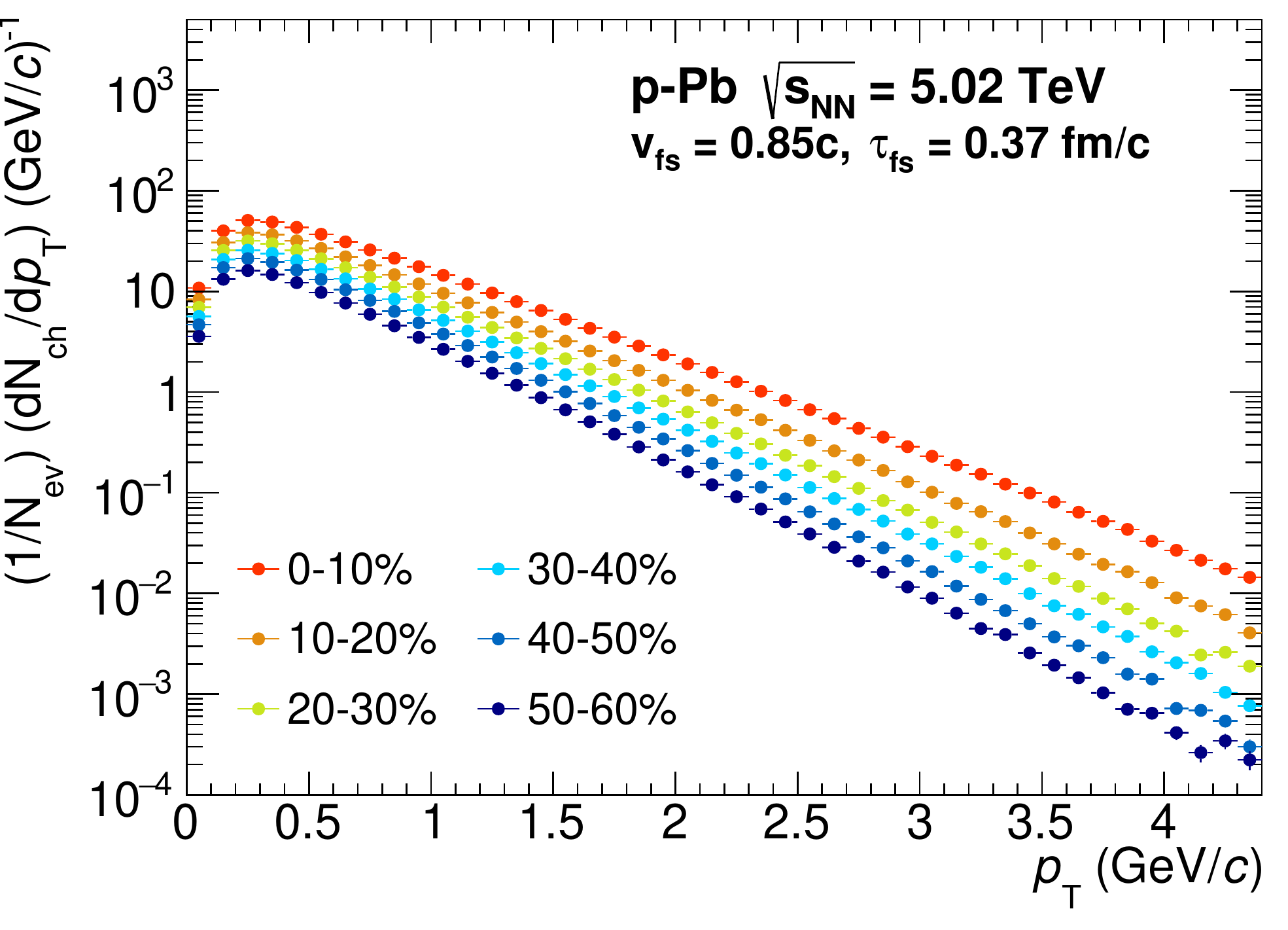}
  \includegraphics[width=.475\linewidth]{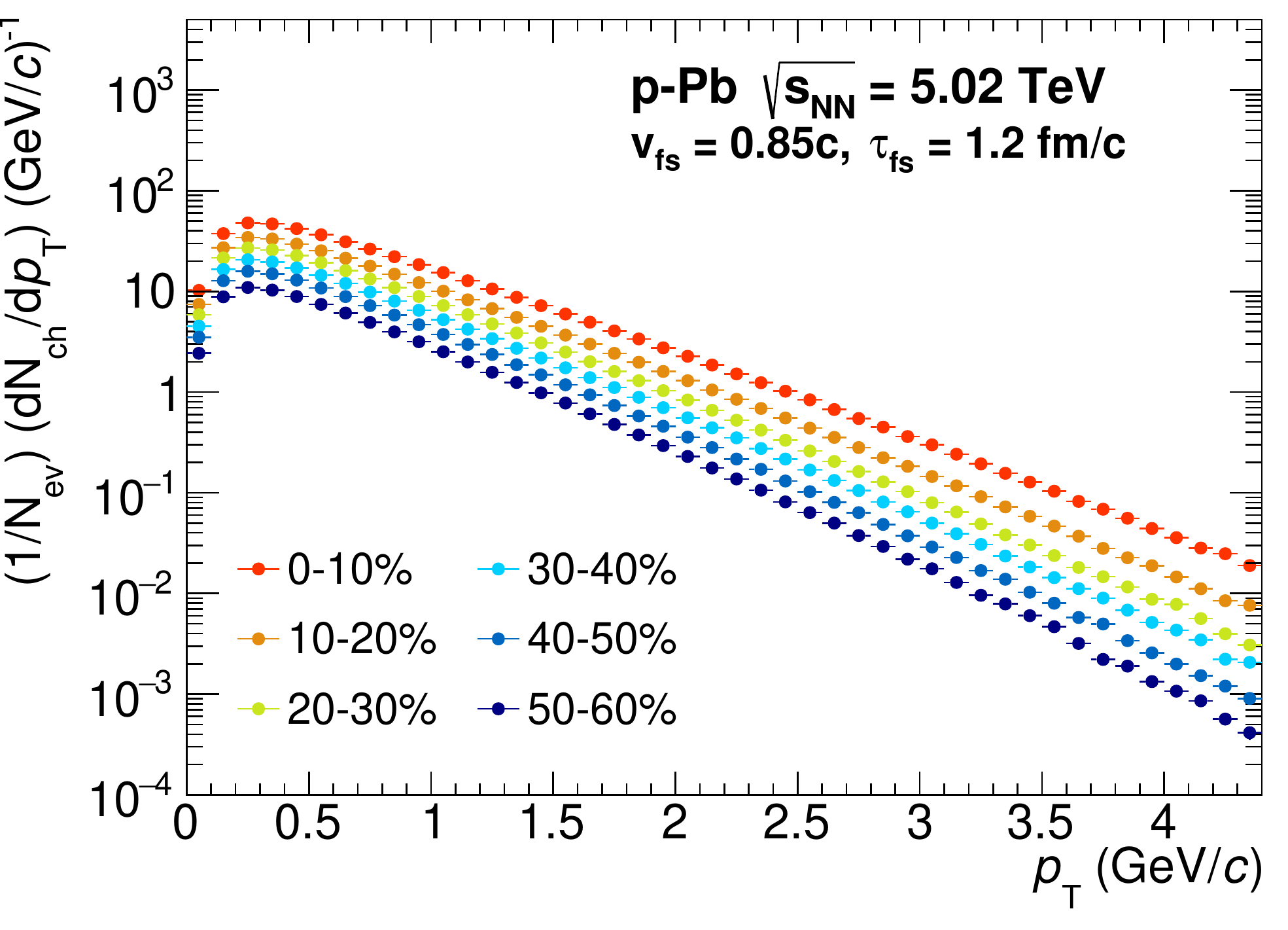}
\caption{Transverse momentum spectra for several centrality classes in p-Pb events for $v_{fs} = c$ and $v_{fs} = 0.85\,c$ and for $\tau_{fs} = 0.37 \text{fm/$c$}$ and $\tau_{fs} = 1.2 \text{fm/$c$}$.}
  \label{fig:pT-Spectra-pPb}
\end{figure}

\begin{figure}[!ht]
  \includegraphics[width=.475\linewidth]{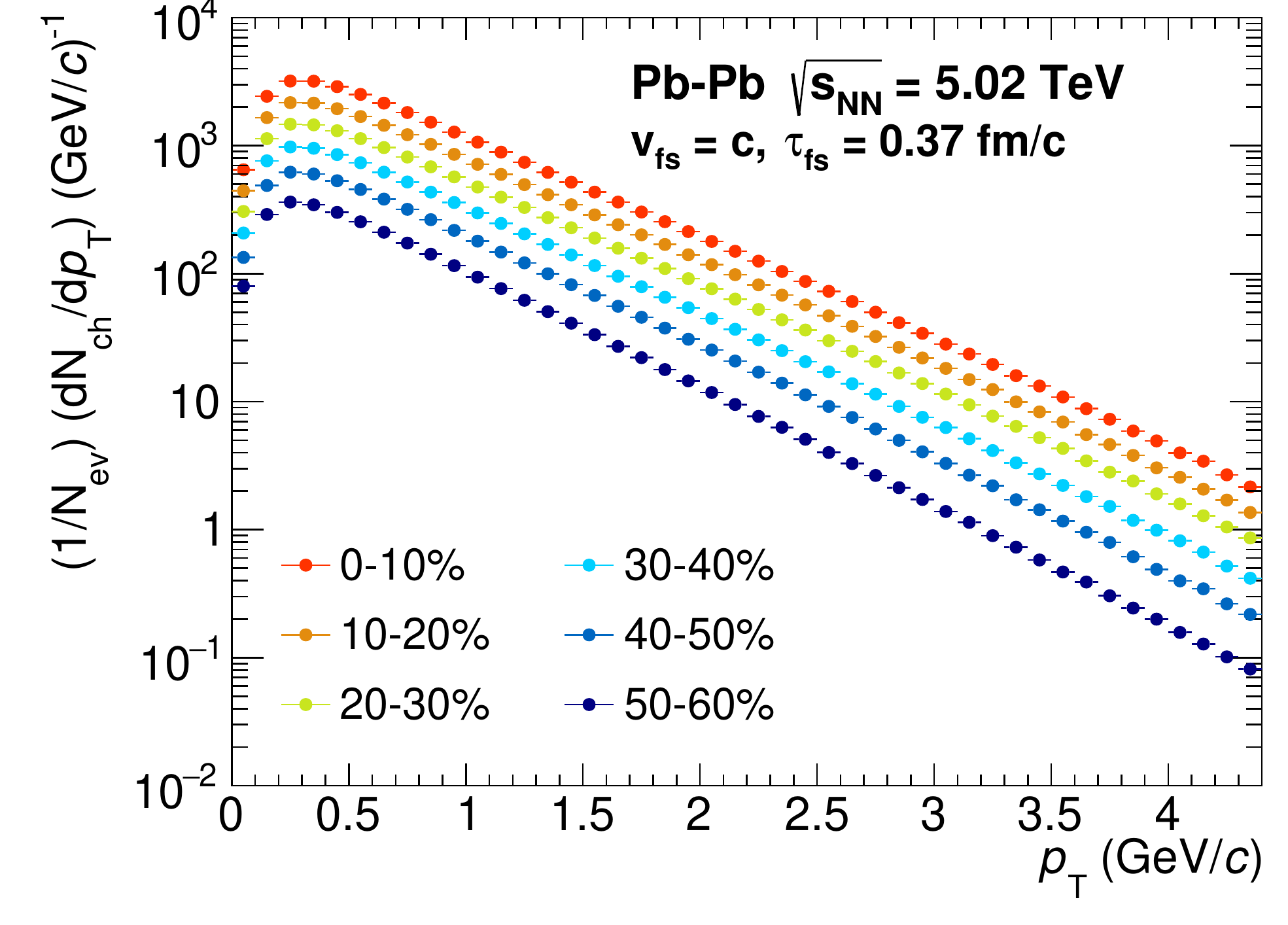}
  \includegraphics[width=.475\linewidth]{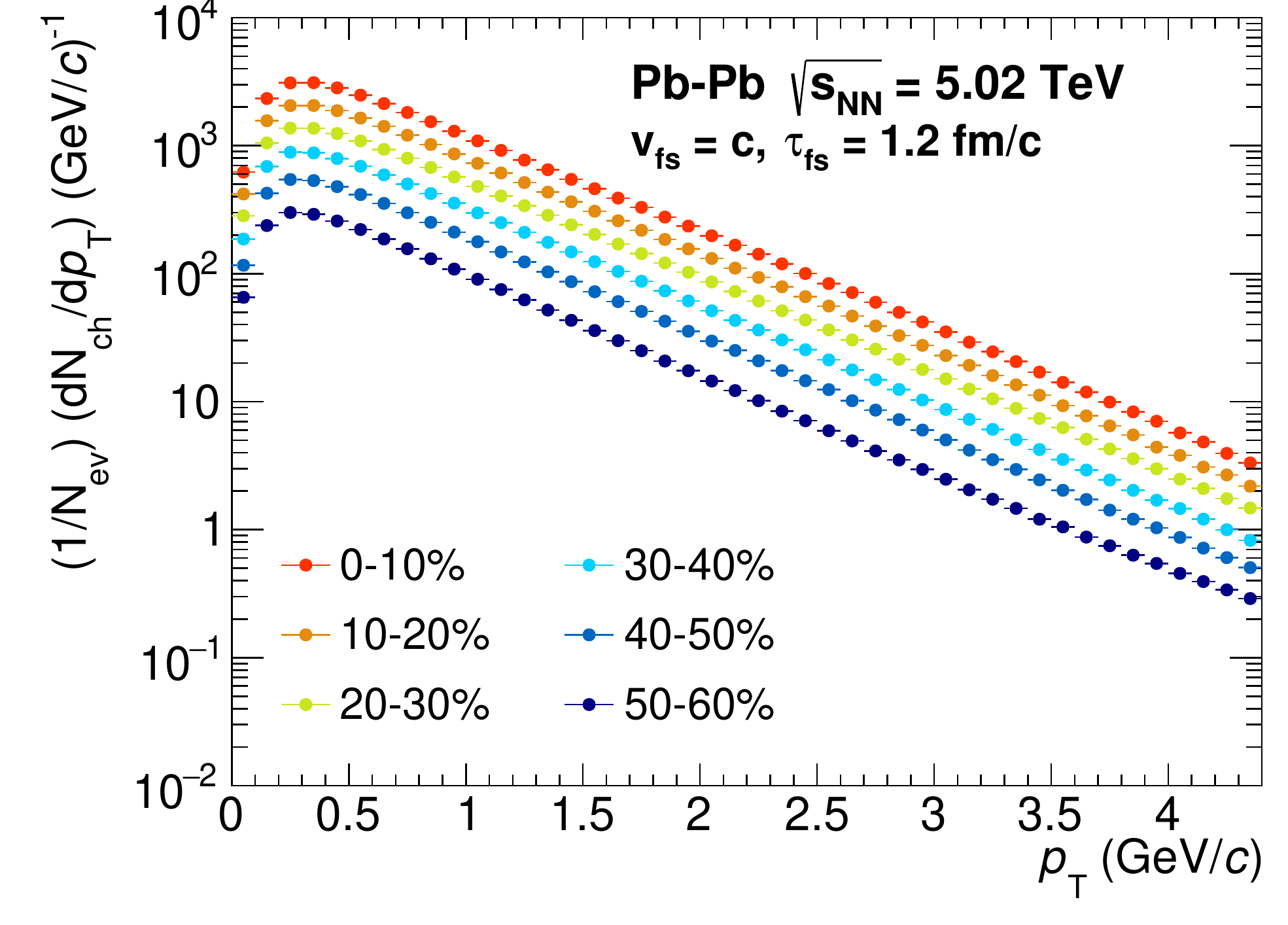}
  \includegraphics[width=.475\linewidth]{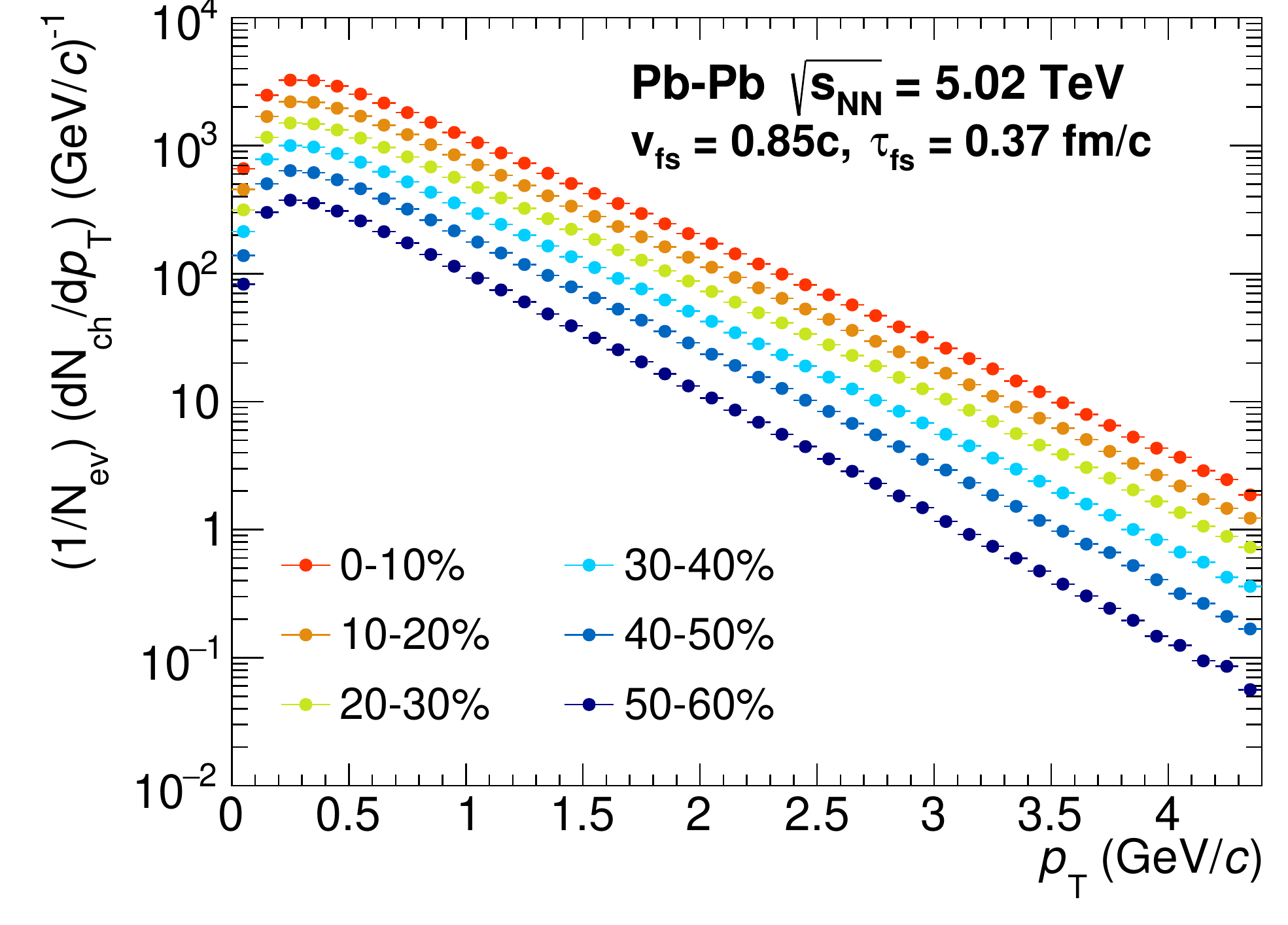}
  \includegraphics[width=.475\linewidth]{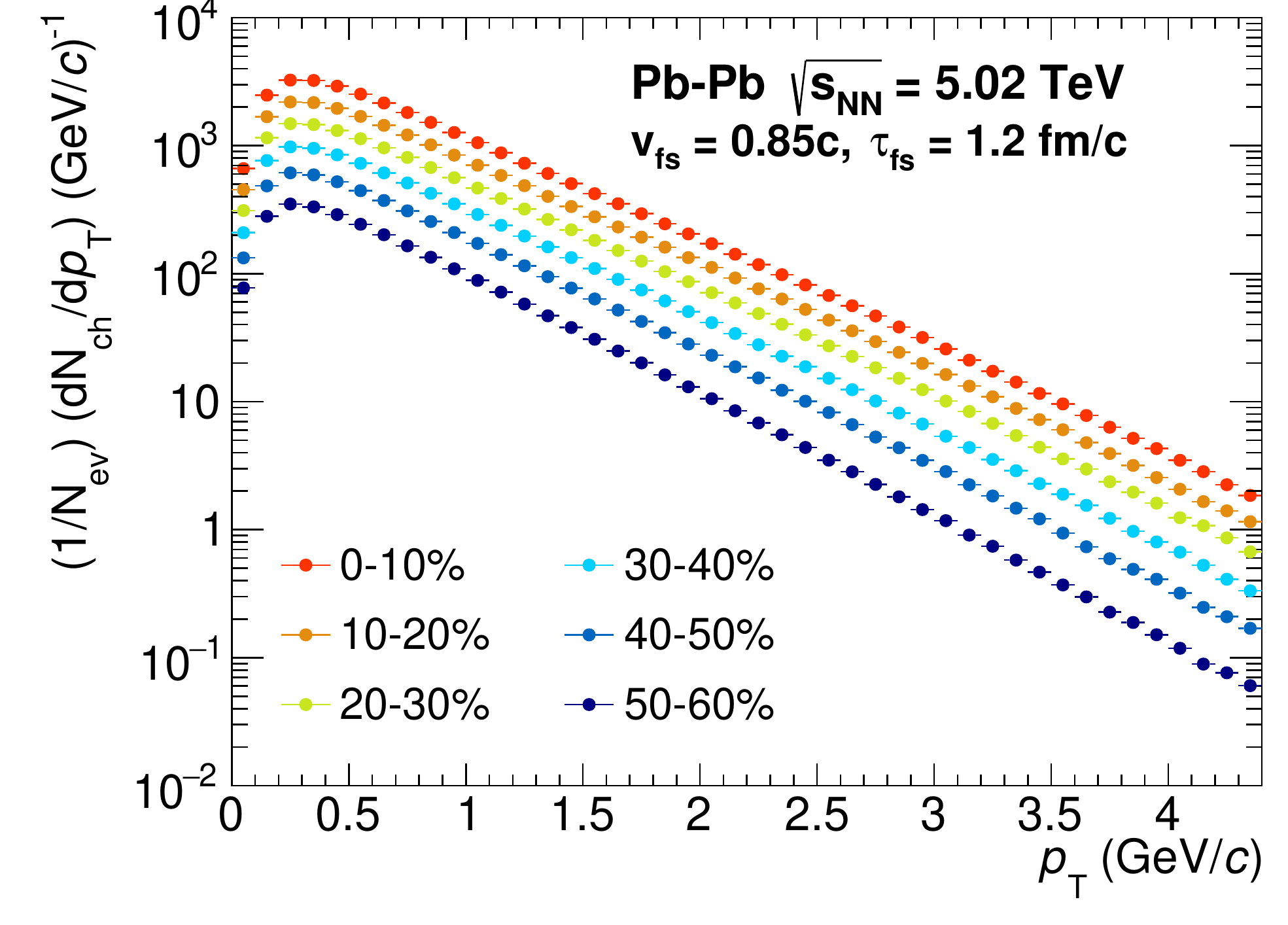}
\caption{Transverse momentum spectra for several centrality classes in Pb-Pb events for $v_{fs} = c$ and $v_{fs} = 0.85\,c$ and for $\tau_{fs} = 0.37 \text{ fm/$c$}$ and $\tau_{fs} = 1.2 \text{ fm/$c$}$.}
  \label{fig:pT-Spectra-PbPb}
\end{figure}

\begin{figure}[!ht]
  \includegraphics[width=.475\linewidth]{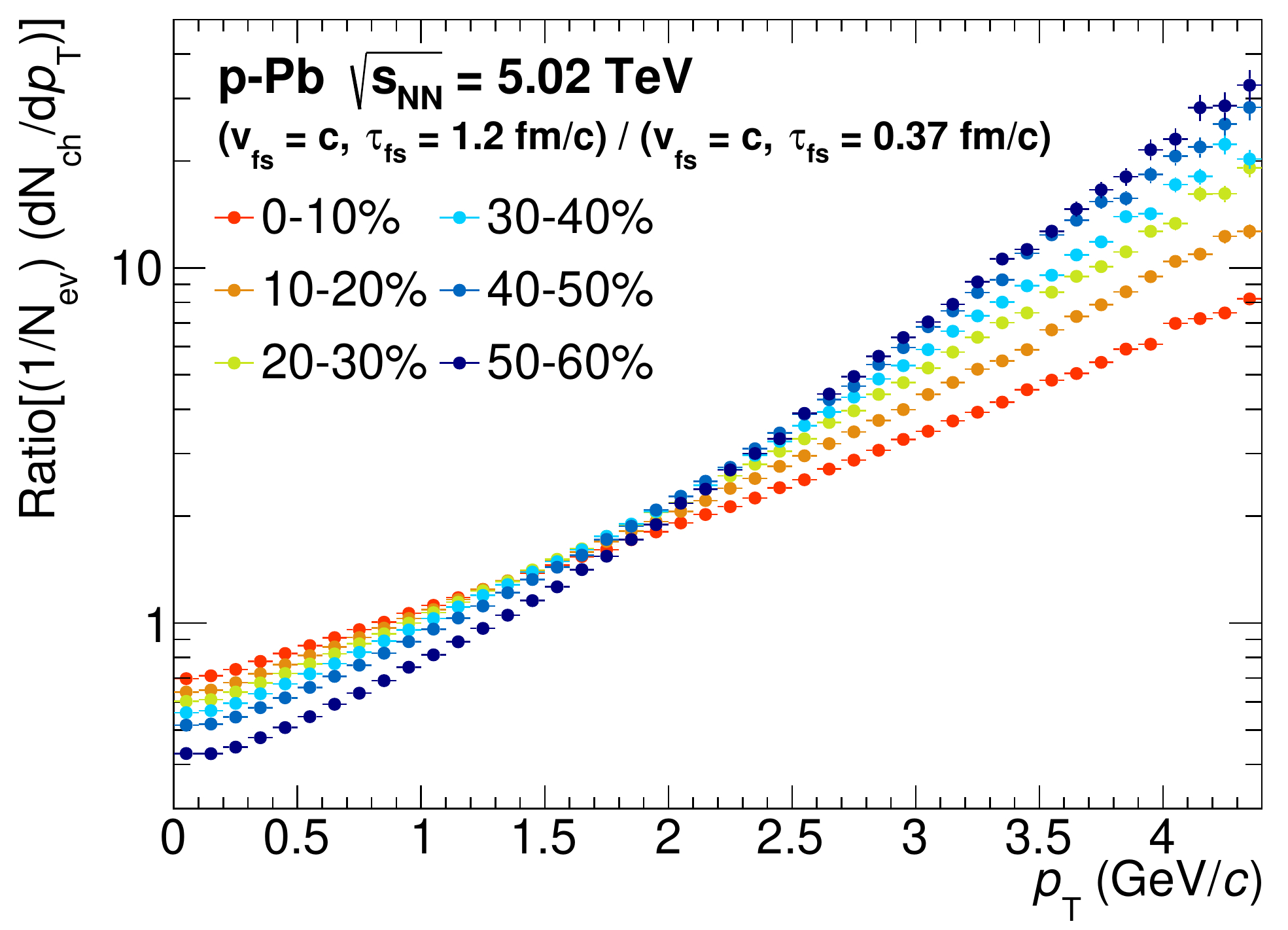}
  \includegraphics[width=.475\linewidth]{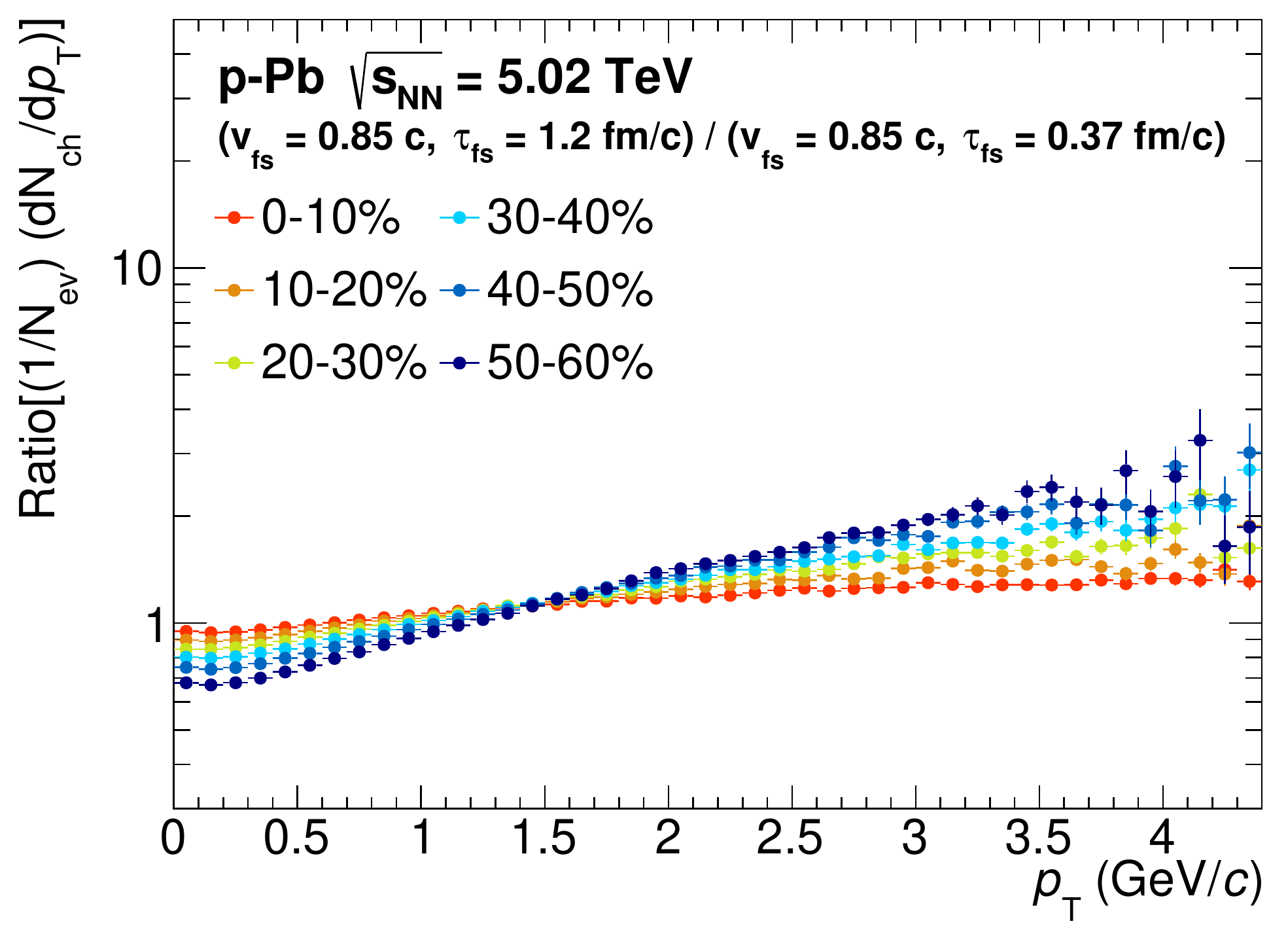}
  \caption{Transverse momentum spectra ratios: for both $v_{fs} = 0.85\,c$ and $v_{fs} = c$, we present the ratio between the spectra obtained with $\tau_{fs} = 1.2$ fm/$c$ and $\tau=0.37$ fm/$c$. Results are presented for several centrality classes p-Pb events.}
  \label{fig:pT-Spectra-ratio-pPb}
\end{figure}

\begin{figure}[!ht]
  \includegraphics[width=.475\linewidth]{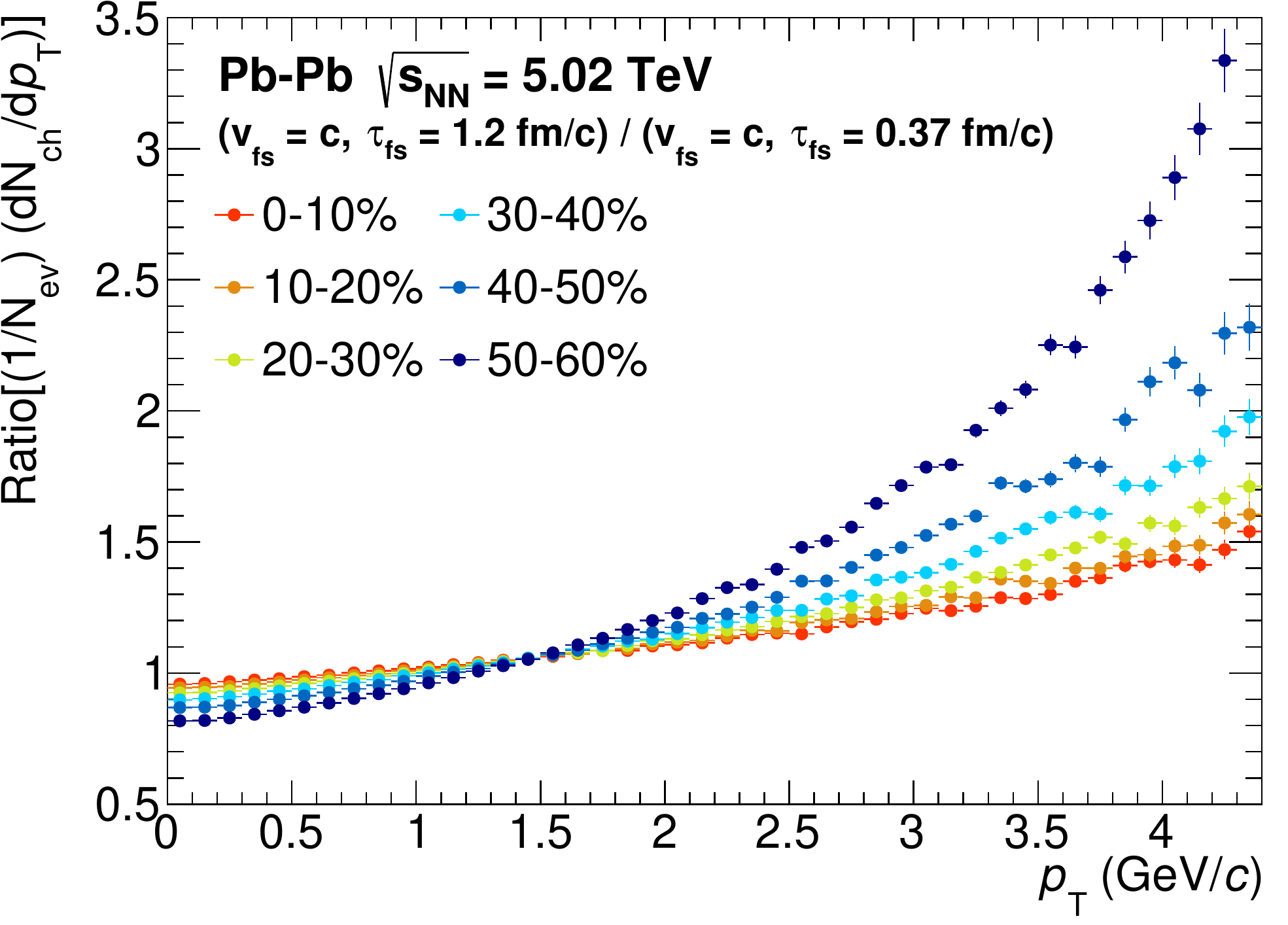}
  \includegraphics[width=.475\linewidth]{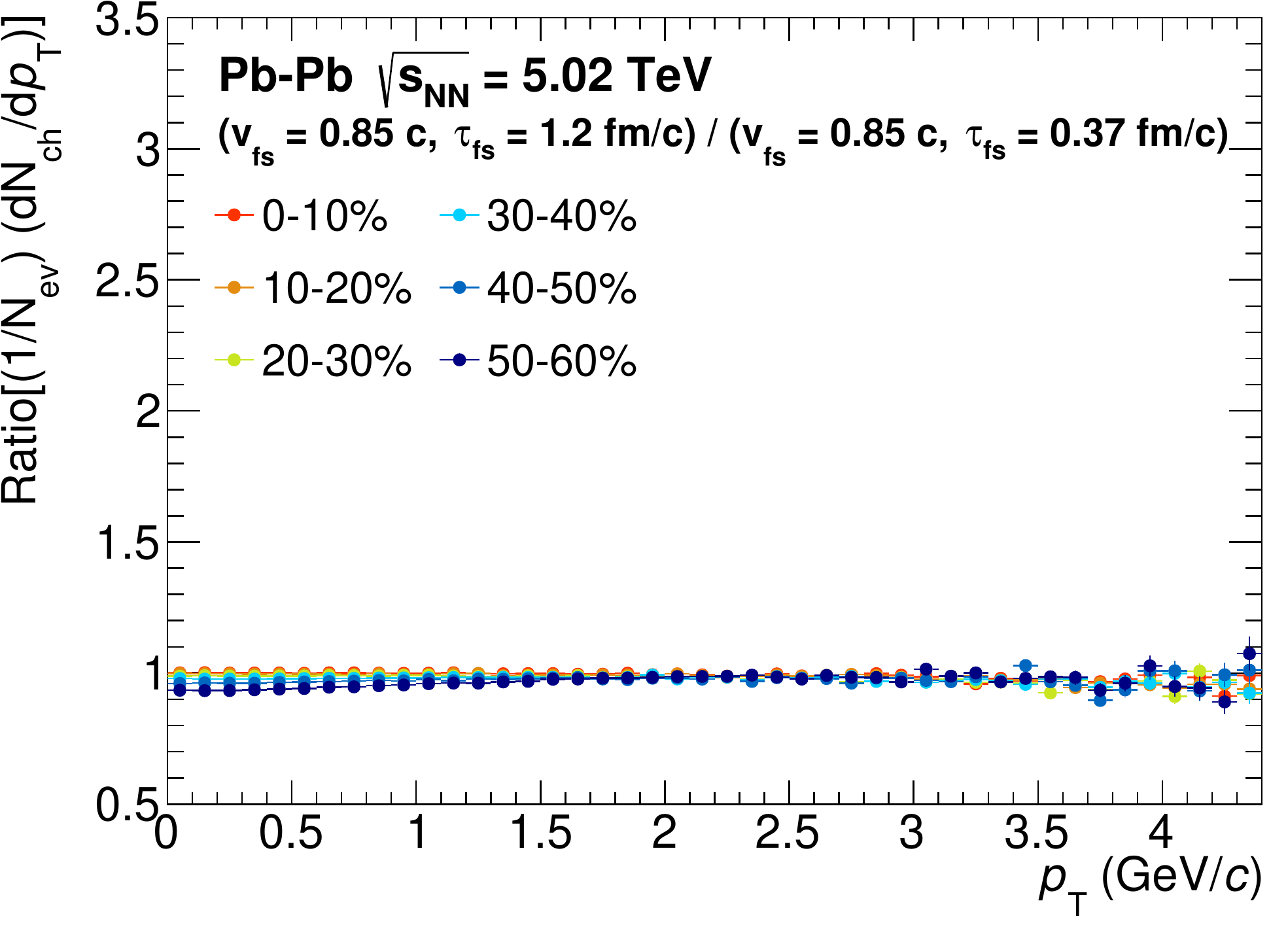}
  \caption{Transverse momentum spectra ratios: for both $v_{fs} = 0.85\,c$ and $v_{fs} = c$, we present the ratio between the spectra obtained with $\tau_{fs} = 1.2$ fm/$c$ and $\tau=0.37$ fm/$c$. Results are presented for several centrality classes Pb-Pb events.}
  \label{fig:pT-Spectra-ratio-PbPb}
\end{figure}

As can be seen in Figures~\ref{fig:pT-Spectra-pPb} and \ref{fig:pT-Spectra-PbPb}, the results imply a sizeable change to the transverse momentum spectra amongst the different combinations of $v_{fs}$ and $\tau_{fs}$. For both p-Pb and Pb-Pb events, when comparing events with different values of $\tau_{fs}$ for fixed $v_{fs}$, one notes that the effect is significantly more pronounced in the conformal case. This can be more clearly visualized by plotting the ratios between spectra with different values of $\tau_{fs}$ for a fixed value of $v_{fs}$. This is shown in Figures~\ref{fig:pT-Spectra-ratio-pPb} and \ref{fig:pT-Spectra-ratio-PbPb}.

Similarly to what was previously observed in Pb-Pb collisions \cite{NunesdaSilva:2020bfs}, for larger free-streaming time, there is less multiplicity in the low-$p_T$ range and, gradually, the trend is inverted around the $p_T$ range $1.5 - 2.0$ GeV. This effect is also larger for peripheral events in comparison to more central ones. It grows when $\tau_{fs}$ is increased, clearly exhibiting the dependence of the magnitude of the effect with the free-streaming duration.

\begin{figure}[!ht]
  \includegraphics[width=.475\linewidth]{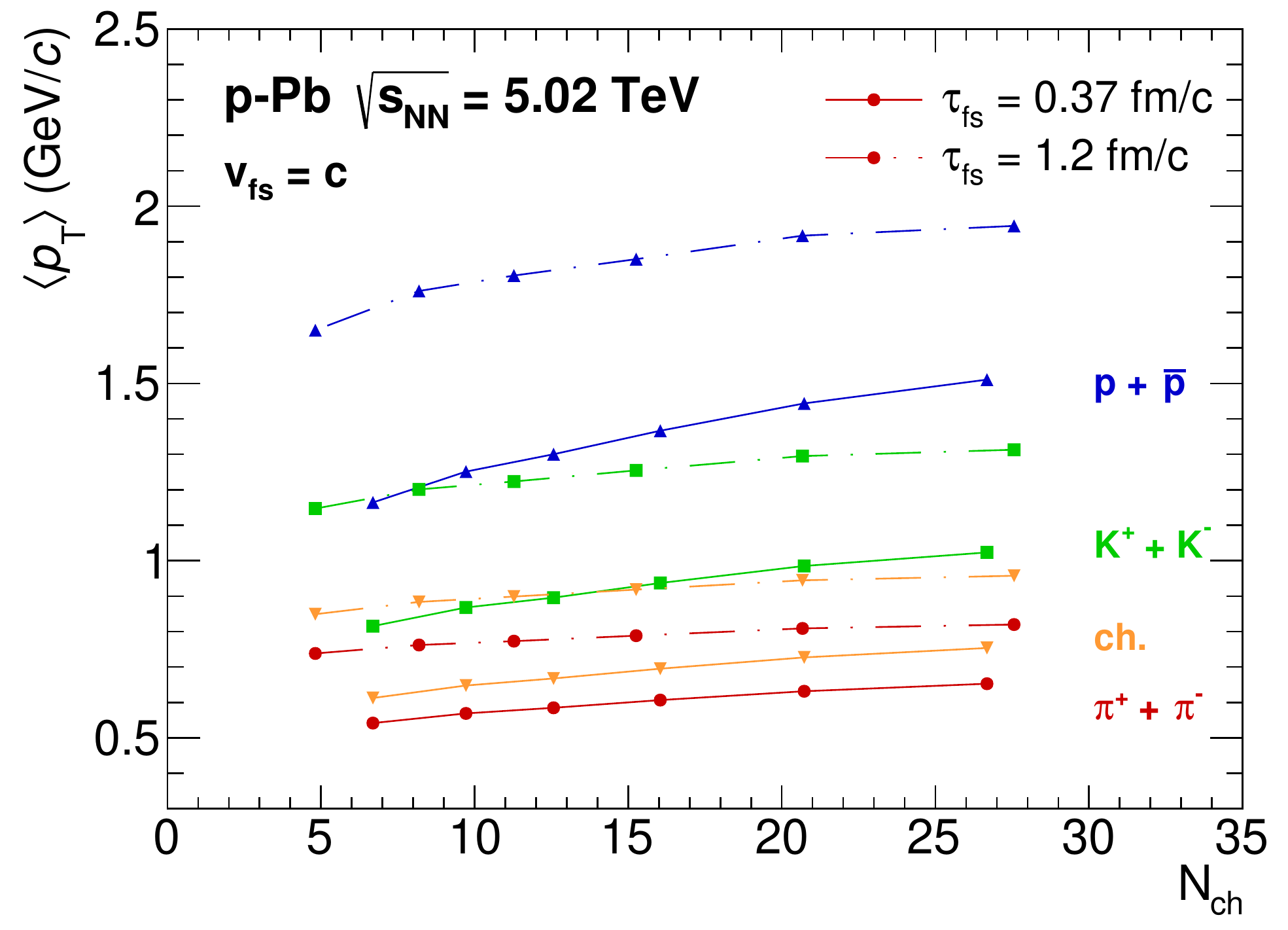}
  \includegraphics[width=.475\linewidth]{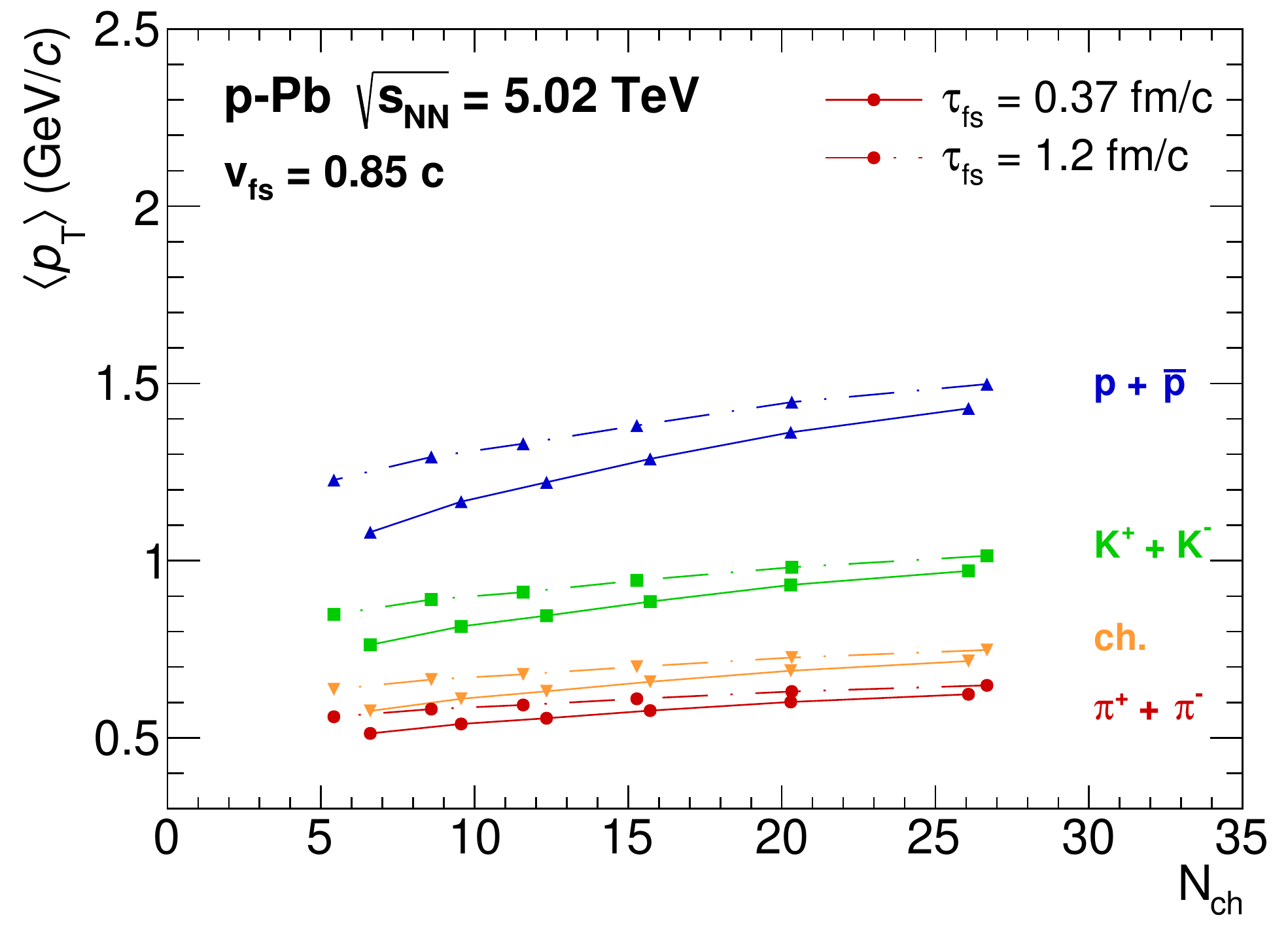}
  \caption{Mean transverse momentum of charged and identified particles for the different scenarios under investigation. Results are presented as a function of charged-particle multiplicity for p-Pb events.}
  \label{fig:mean-pT}
\end{figure}

\begin{figure}[!ht]
  \includegraphics[width=.475\linewidth]{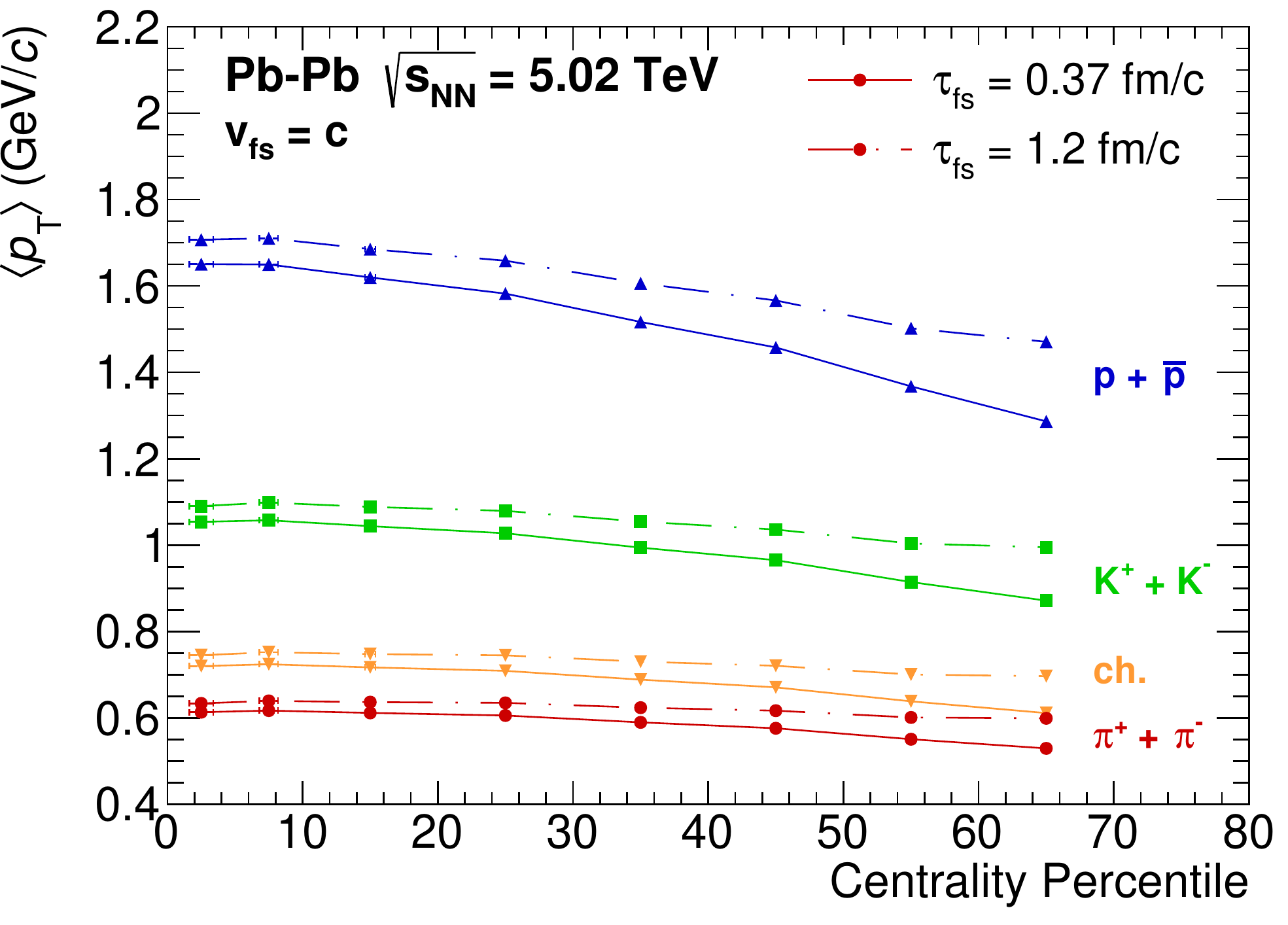}
  \includegraphics[width=.475\linewidth]{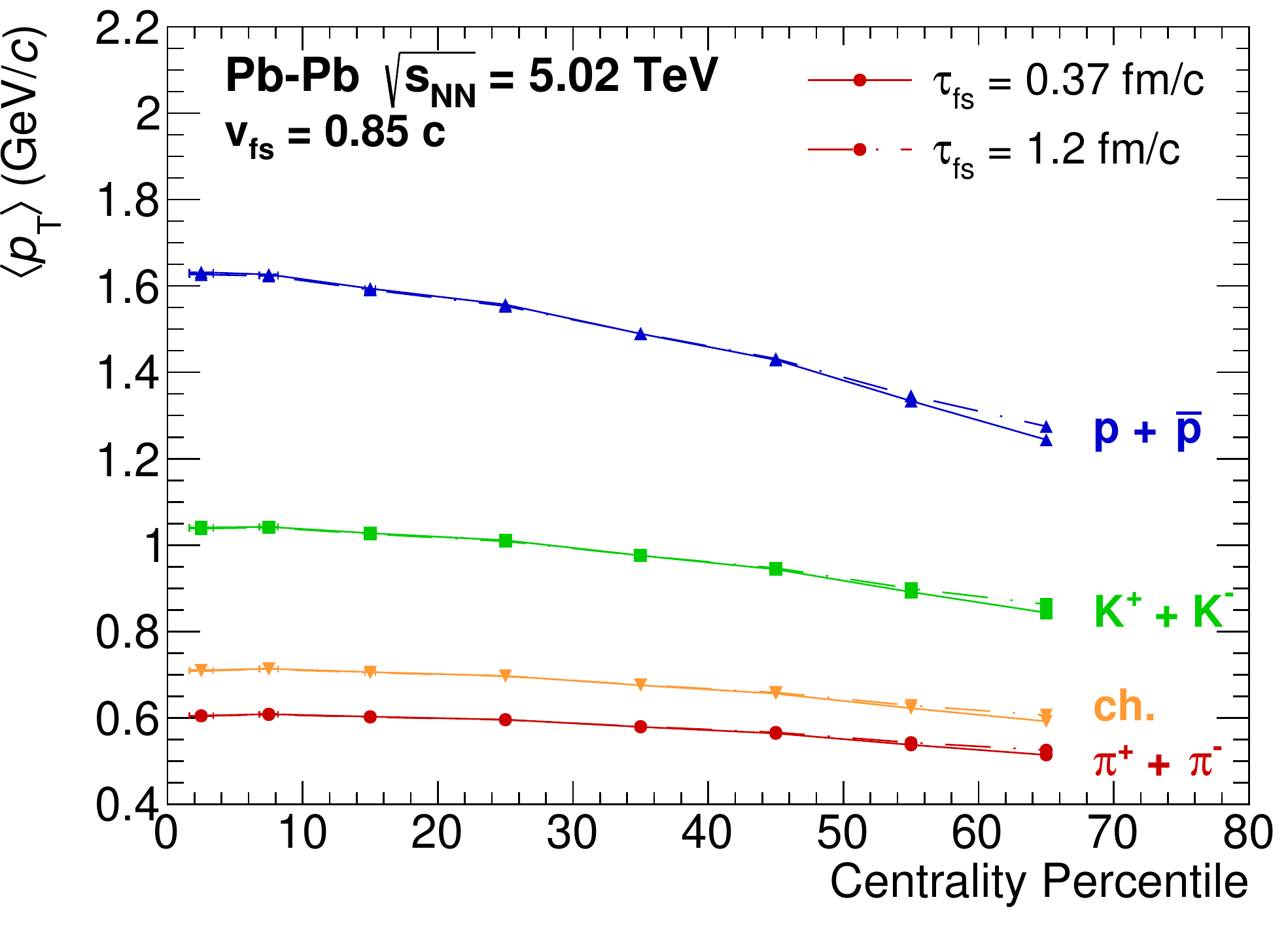}
  \caption{Mean transverse momentum of charged and identified particles for the different scenarios under investigation. Results are presented as a function of centrality for Pb-Pb events.}
  \label{fig:mean-pT_PbPb}
\end{figure}

These effects can be also seen when looking into the mean transverse momentum. Results for the mean transverse momentum of charged and some identified particles, in the different scenarios under investigation, are presented in Figures~\ref{fig:mean-pT} and~\ref{fig:mean-pT_PbPb}. 
These results again indicate that the longer the duration of the pre-equilibrium stage, the more transverse momentum is artificially added into the system. Note, however, that the magnitude of this effect is significantly reduced in the non-conformal scenario, in comparison to the conformal case. 

\begin{figure}[!ht]
  \includegraphics[width=.475\linewidth]{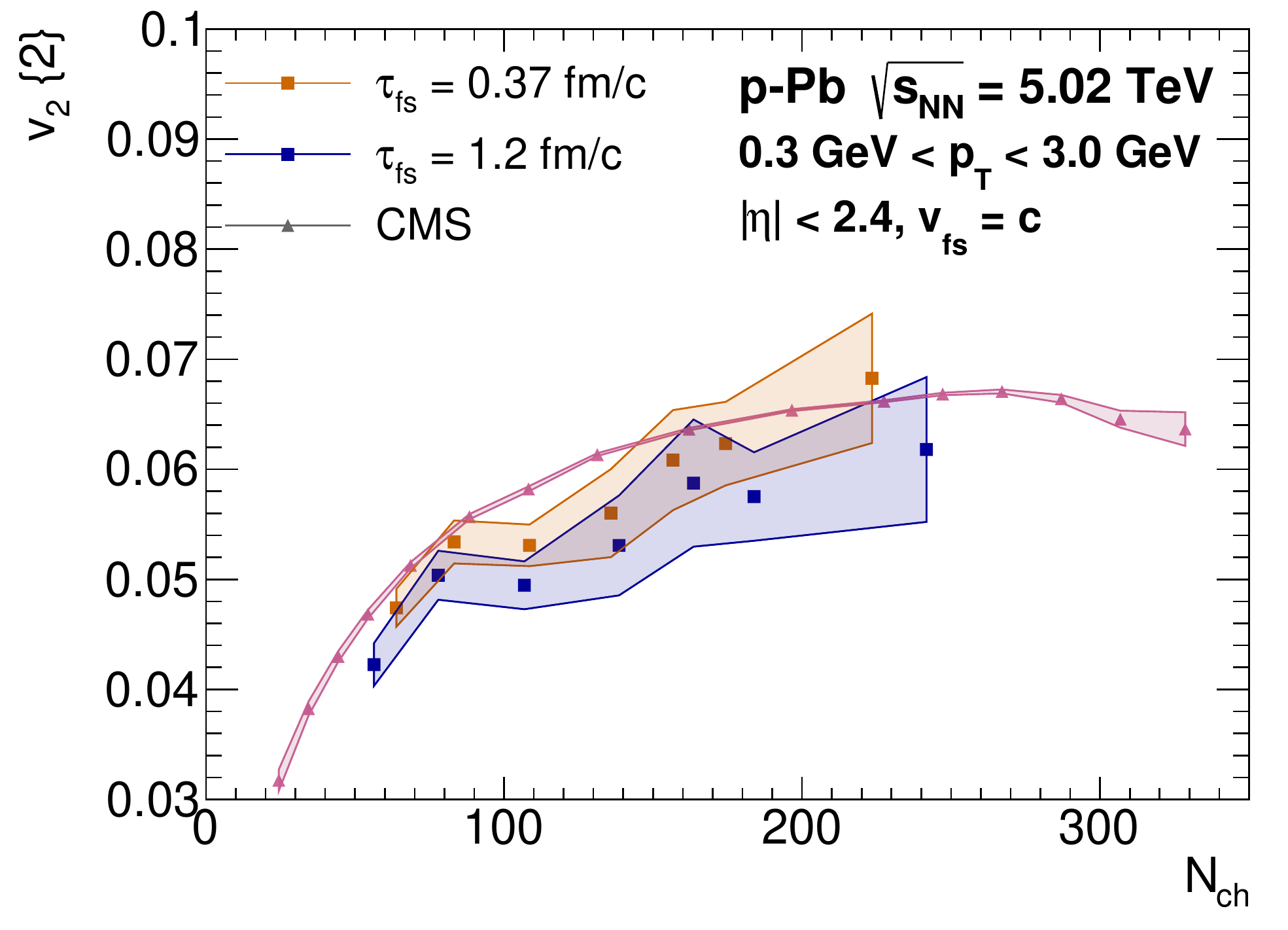}
  \includegraphics[width=.475\linewidth]{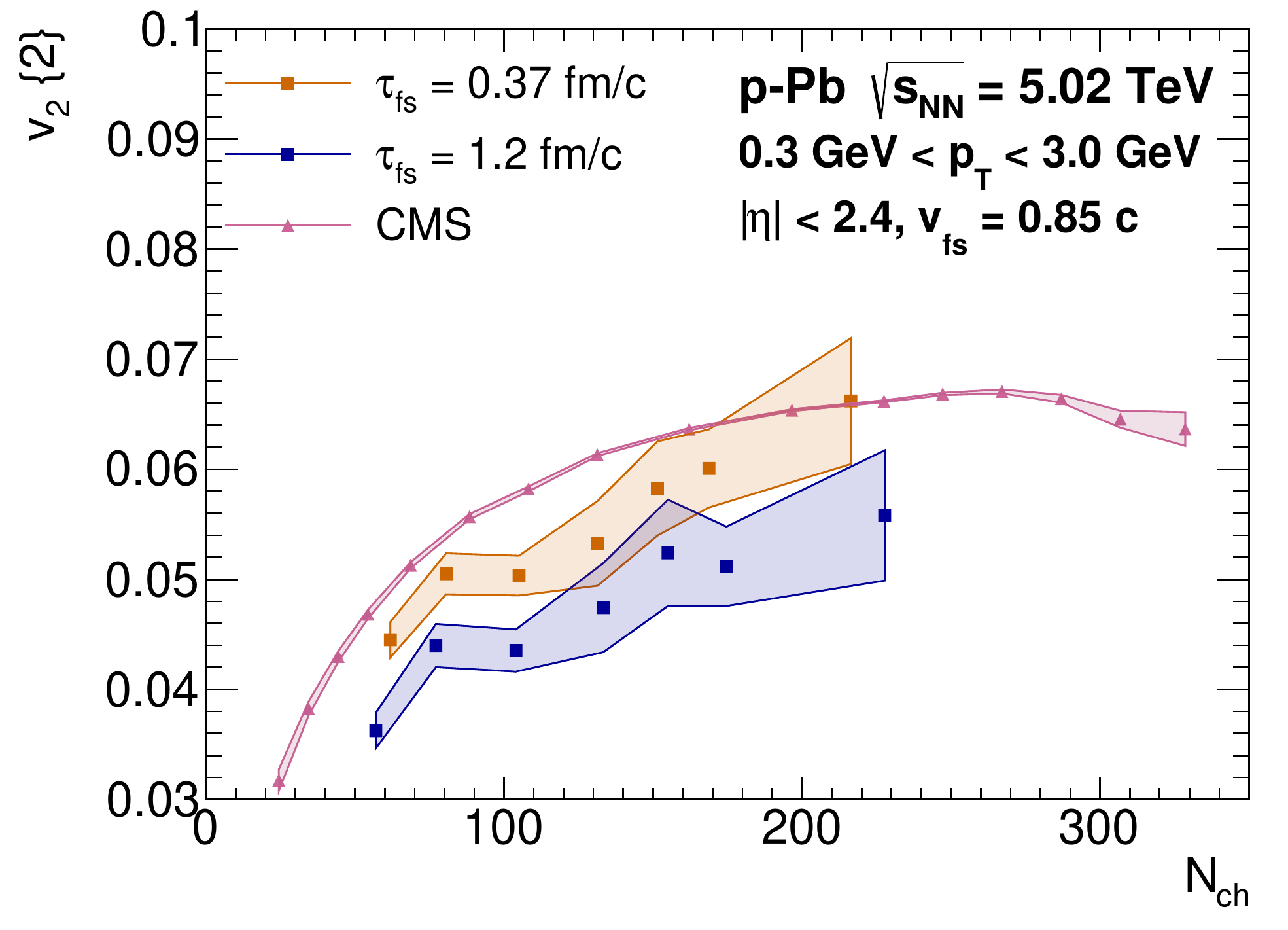}
  \caption{Anisotropic flow coefficient from two-particle correlations $v_2\{2\}$ for the p-Pb scenarios under investigation plotted as a function of charge particle multiplicity. Experimental data from the CMS collaboration were obtained from Ref.~\cite{CMS:2013jlh}}
  \label{fig:v22-pPb}
\end{figure}

\begin{figure}[!ht]
  \includegraphics[width=.475\linewidth]{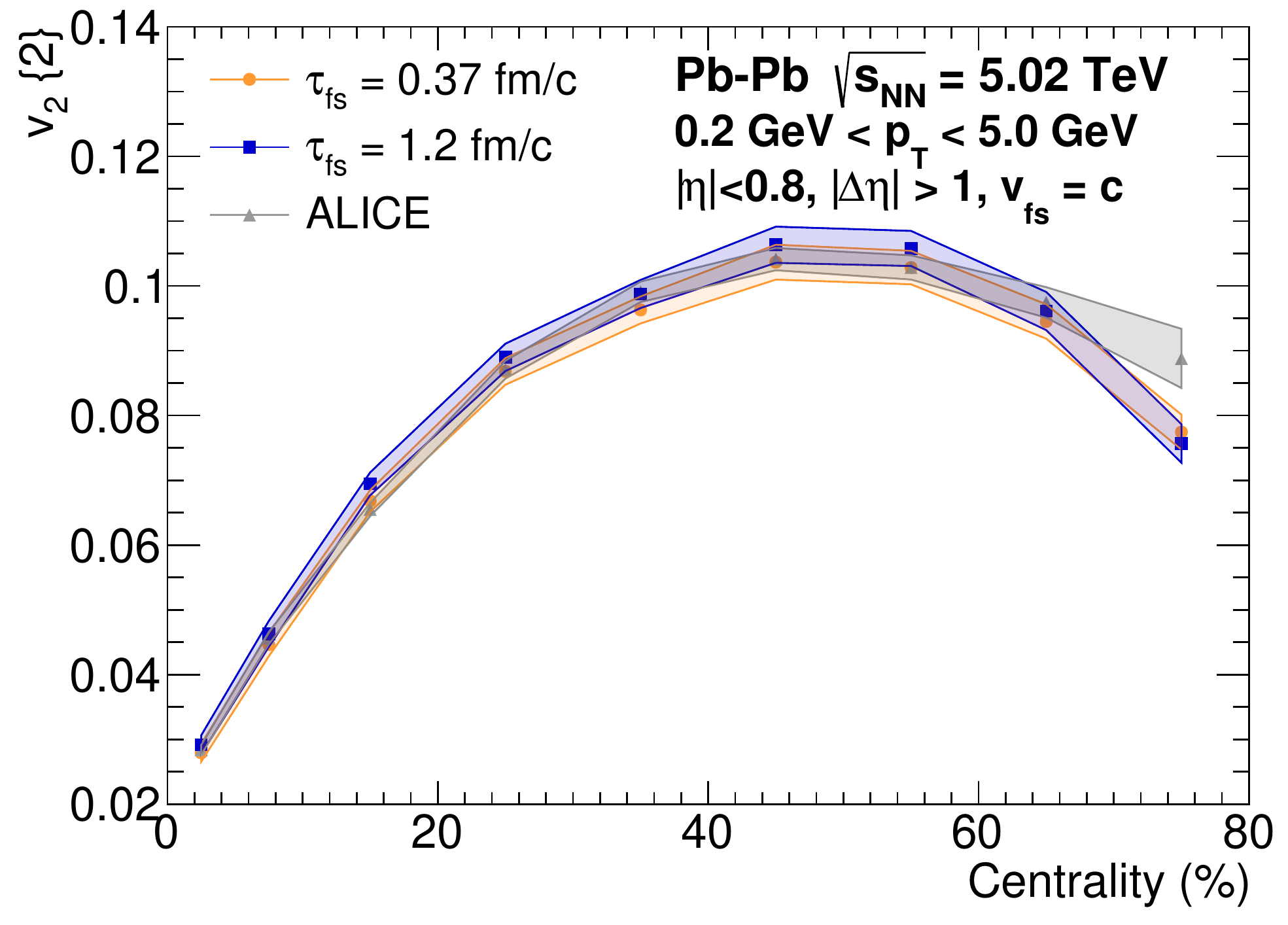}
  \includegraphics[width=.475\linewidth]{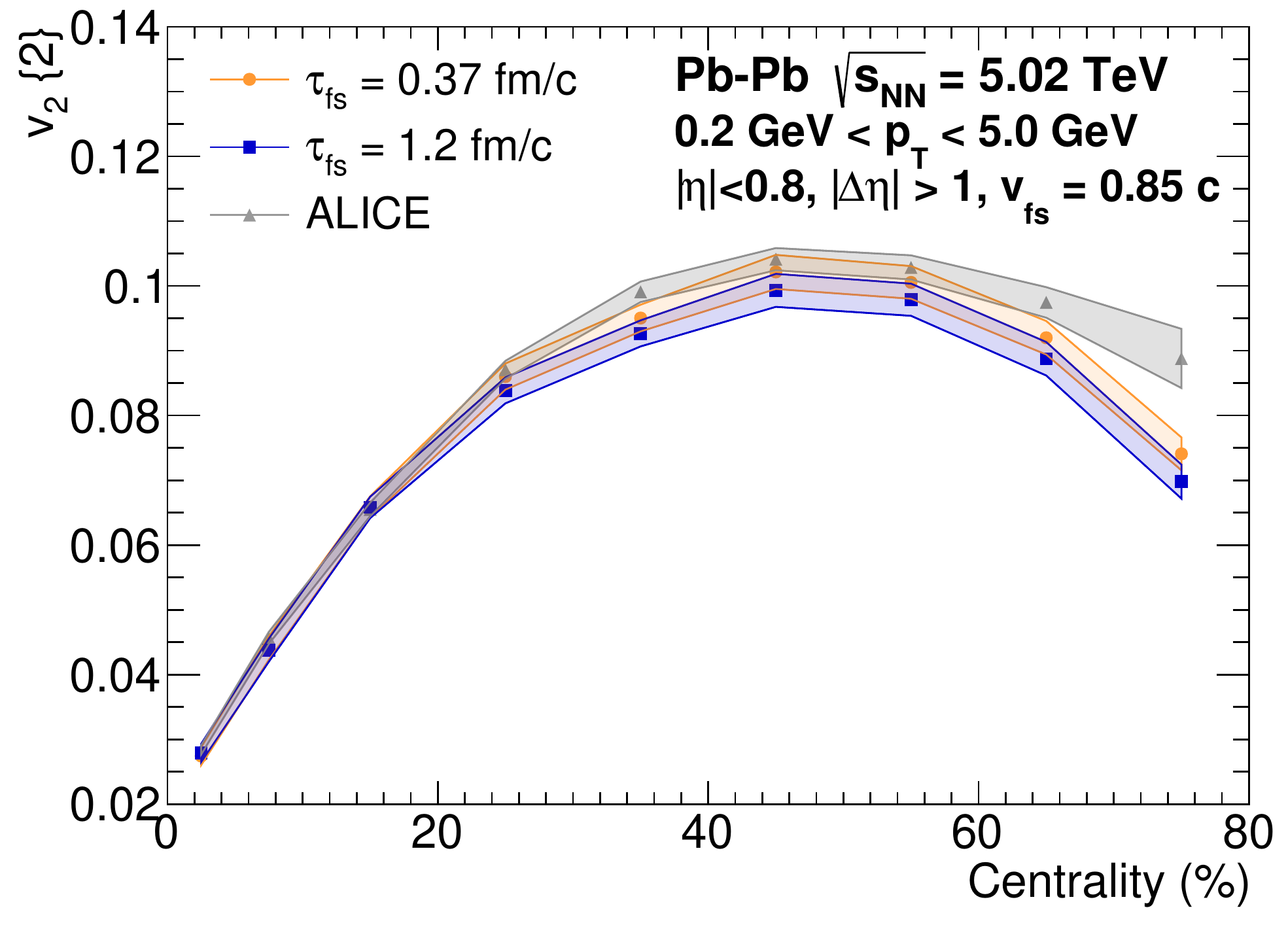}
  \caption{Anisotropic flow coefficient from two-particle correlations $v_2\{2\}$ for the Pb-Pb scenarios under investigation plotted as a function of event centrality. Experimental data from the ALICE collaboration were obtained from Ref.~\cite{ALICE:2016ccg}.}
  \label{fig:v22-PbPb}
\end{figure}

These modifications to the transverse momentum spectrum should also contaminate other $p_T$-integrated observables. One example is the usual anisotropic flow coefficients in the Fourier expansion of the transverse momentum probability distribution at a certain azimuthal angle $\phi$, 
\begin{equation}
E\frac{dN}{d^3 p} \equiv \frac{1}{2\pi} \frac{\sqrt{m^2 + p_T^2 \cosh{(\eta)^2}}}{p_T \cosh{(\eta)}} \frac{dN}{p_T dp_T d\eta} \left[ 1 + \sum_{n=1}^\infty v_n(p_T, y) \cos{n(\phi - \Psi_n)} \right].
\end{equation}
We have calculated the $v_2\{2\}$ observable for the scenarios under investigation, employing the Q-cumulants formalism, thus avoiding event plane calculations. The two-particle flow coefficients are connected to the two-particle correlation function through the relation \cite{Borghini:2001vi, Bilandzic:2010jr}
\begin{equation}
     v_n\{2\} = \sqrt{\langle v_n \rangle^2} = \sqrt {\langle\langle e^{in(\phi_1 - \phi_2)}\rangle\rangle}.
\end{equation}
By calculating the so-called Q-vector \cite{Bilandzic:2010jr} 
\begin{equation}
    Q_n = \sum_{i=1}^M e^{in\phi_i},
\end{equation}
it is possible to extract the $v_n$ coefficients using the relation
\begin{equation}
    v_n\{2\} = \sqrt{ \left\langle \frac{\lvert Q_n \lvert^2 - M}{M(M-1)}\right\rangle},
\end{equation}
where $M$ is the event-by-event multiplicity in the analyzed window. 
The resulting values for $v_2\{2\}$ are presented in Figures~\ref{fig:v22-pPb} and \ref{fig:v22-PbPb} for the different scenarios for both p-Pb and Pb-Pb systems.

It can be seen that the $v_{fs} = c$, $\tau_{fs} = 0.37$ fm/$c$ scenario has a reasonable agreement with experimental data for both system sizes. This is expected since we have used the maximum a posteriori parameters obtained in \cite{Moreland:2018gsh} with $v_{fs}=c$, which yield $\tau_{fs}=0.37$ fm/$c$. Again, the smaller system results are more sensitive to variations of both $\tau_{fs}$ and $v_{fs}$. For both system sizes, at fixed free-streaming velocities, longer free-streaming periods lead to diminishing flow, as a consequence of the changes in the transverse momentum spectra. 

It has been argued in \cite{NunesdaSilva:2020bfs} that a significant part of the momentum that conformal free streaming models add to the system comes from the above-mentioned unphysical enhancement of the initial bulk viscous pressure at the switch to hydrodynamics. We now compare this effect between large and small systems. To this goal, we have simulated additional sets of events, corresponding to the scenarios under investigation, but ignoring the initial bulk pressure at the switch to hydrodynamics, thus effectively removing the artificial bulk pressure that appears in conformal free-streaming models. 

\begin{figure}[!ht]
  \includegraphics[width=.475\linewidth]{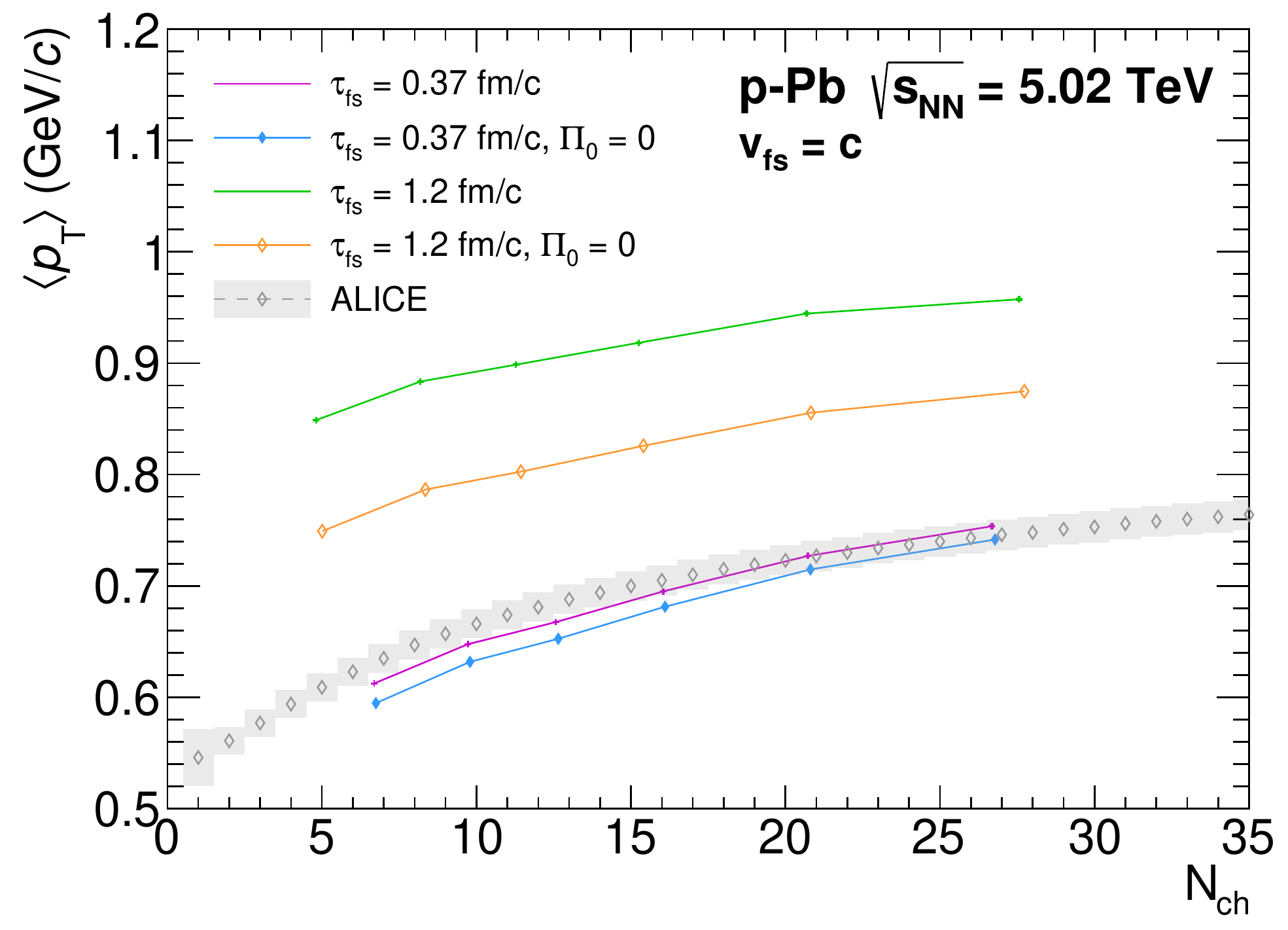}
  \includegraphics[width=.475\linewidth]{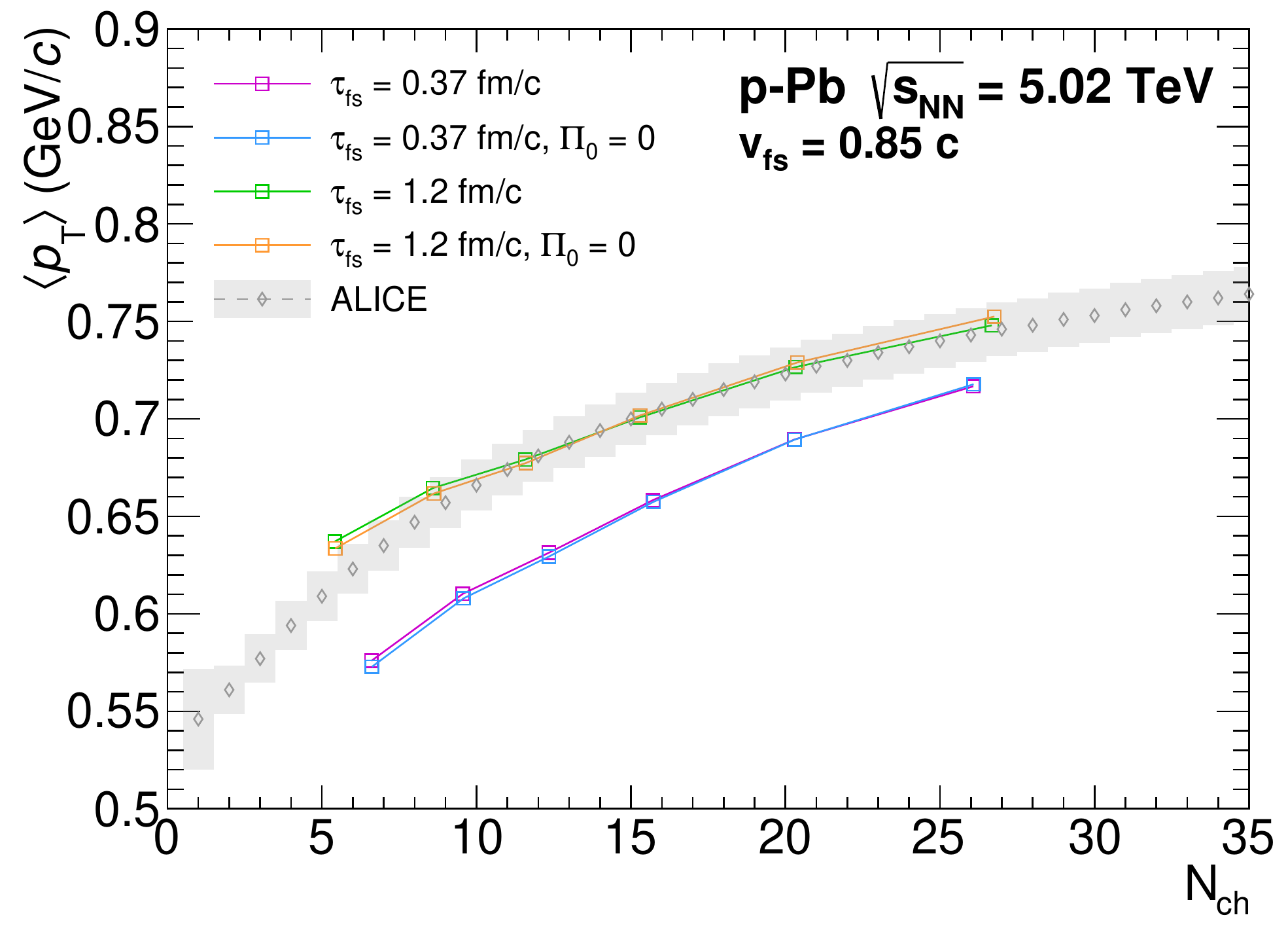}
  \caption{Mean transverse momentum of charged particles for the different scenarios under investigation. Results are presented as a function of charged-particle multiplicity for p-Pb events. Experimental data from the ALICE collaboration were obtained from Ref.~\cite{ALICE:2013rdo}}
  \label{fig:mean-pT-zeroBulk}
\end{figure}

\begin{figure}[!ht]
  \includegraphics[width=.475\linewidth]{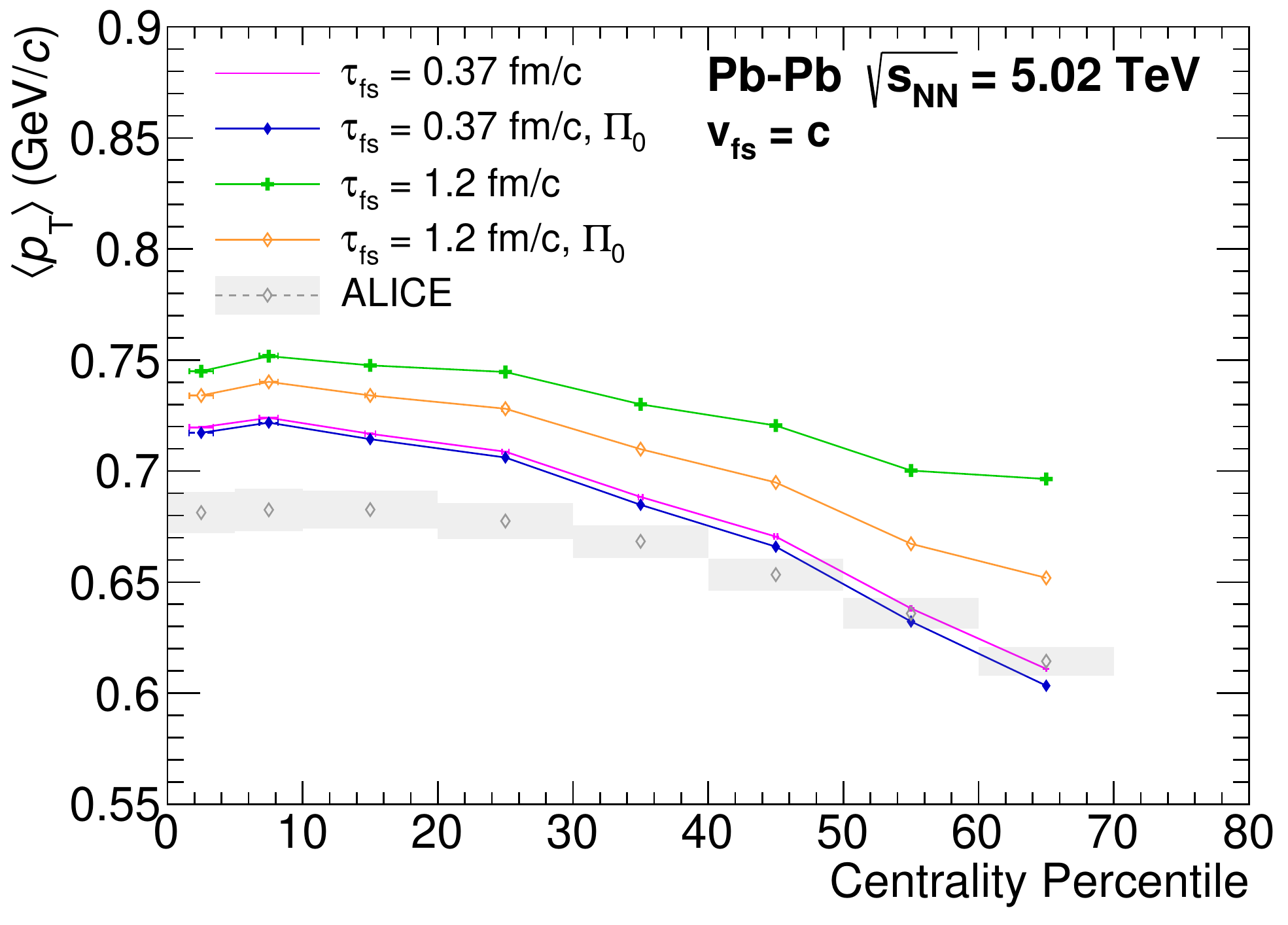}
  \includegraphics[width=.475\linewidth]{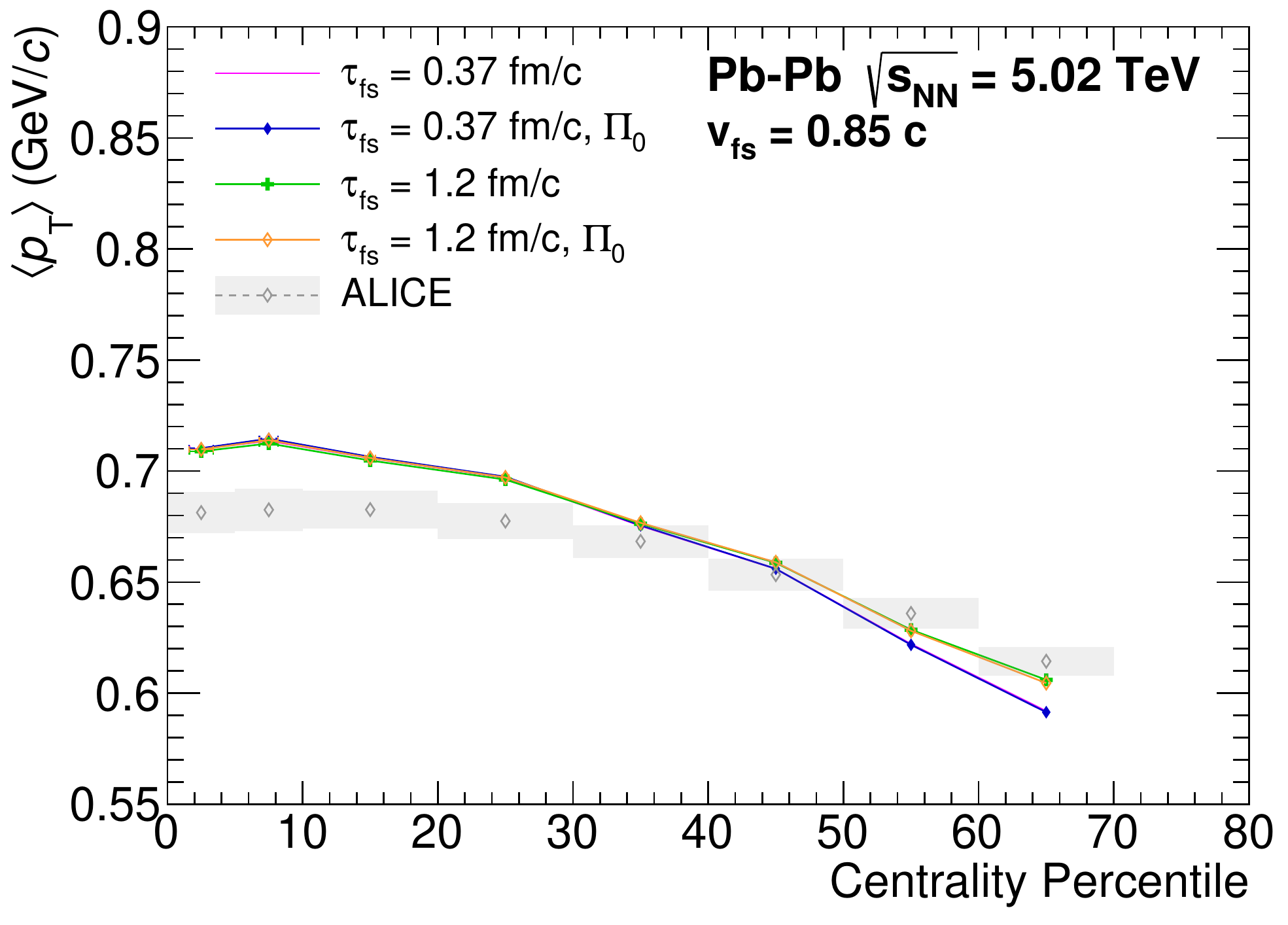}
  \caption{Mean transverse momentum for the different scenarios under investigation. Results are presented as a function of centrality for Pb-Pb events. Experimental data from the ALICE collaboration were obtained from Ref.~\cite{ALICE:2018hza}.}
  \label{fig:mean-pT-zeroBulk_PbPb}
\end{figure}

In Figures~\ref{fig:mean-pT-zeroBulk} and \ref{fig:mean-pT-zeroBulk_PbPb}, we present the mean transverse momentum results for charged particles for our original scenarios in comparison to the equivalent scenarios with $\Pi_0 = 0$.
It is noticeable that the effect observed originally in \cite{NunesdaSilva:2020bfs} is greatly enhanced when one moves to small systems, possibly due to the shorter lifetime of the QGP in such systems. The effect also grows with longer free-streaming durations, as had already been hinted by the larger values of mean transverse momentum noted above (a large fraction of this extra momentum is indeed coming from a such artificial enhancement of the initial bulk viscous pressure). 

Finally, it can be noted that the effect diminishes in the non-conformal scenarios: for $v_{fs}=0.85\,c$, the differences in $\langle p_T\rangle$ when $\Pi_0$ is set to zero is significantly smaller than in the conformal case, even for p-Pb events. This is an important result: it provides further evidence that the artificial enhancement of the initial pressure indeed contaminates final state observables and it also suggests that effectively breaking the conformal symmetry in free-streaming models, even if crudely via a reduced velocity, can alleviate the contamination of final state observables. 

\section{Conclusions}
 In this work, we have extended our analysis of the effect of pre-equilibrium dynamics on final state observables to smaller system sizes, specifically to p-Pb systems. We have also tested whether a subluminal free-streaming pre-equilibrium model can alleviate the effects of the artificial bulk pressure present in most pre-equilibrium scenarios.

We have found that the shorter duration of the hydrodynamical stage within a hybrid model leads to a larger sensibility of final state observables to pre-hydrodynamical physics in smaller systems. All of the effects previously exhibited in Pb-Pb collisions \cite{NunesdaSilva:2020bfs} and also in our new set of Pb-Pb collisions are enhanced in p-Pb simulations, making it clear that the case for moving beyond conformal pre-hydrodynamical frameworks is even more important as we try to understand the formation of QGP in such systems.

We have also seen that the issues related to the unphysical enhancement of the initial bulk viscous pressure are indeed diminished with the use of a non-conformal pre-equilibrium model. This was done by choosing a free-streaming velocity smaller than $c$. While this prescription is certainly very simplistic  (at best), it nevertheless captures the essence of the effect, by effectively removing conformal invariance. More realistic descriptions of the non-conformal pre-equilibrium phase are needed to fully remove, in a consistent way, the problem discussed in this work.      

Our results show that one must still improve the description of the hydrodynamization of QCD matter and its matching to the hydrodynamical regime. Until a more refined microscopic model is available, the artificial enhancement of the bulk pressure at the transition to hydrodynamics should be considered as a caveat in the extraction of transport coefficients from Bayesian studies, especially for small systems. 

\begin{acknowledgments}
This research was funded by FAPESP grants number 2016/13803-2 (D.D.C.), 2021/04924-9 (A.V.G.), 2016/24029-6, 2018/24720-6 (M.L.), 2017/05685-2 (all) and 2018/07833-1 (M.H.). D.D.C., M.L., G.S.D., and J.T. thank CNPq for their financial support. T.N.dS. acknowledges financial support from CNPq, grant number 409029/2021-1. M.H. was supported in part by the National Science Foundation (NSF) within the framework of the MUSES collaboration, under grant number OAC-2103680. G.S.D. acknowledges financial support from Funda\c c\~ao Carlos Chagas Filho de Amparo \`a Pesquisa do Estado do Rio de Janeiro (FAPERJ), grant number E-26/202.747/2018. J.N. is partially supported by the U.S. Department of Energy, Office of Science, Office for Nuclear Physics under Award No. DE-SC0021301.
\end{acknowledgments}

\bibliography{ref}

\end{document}